\documentclass[useAMS,usenatbib,times]{mn2e}

\usepackage[]{graphicx}
\usepackage{graphicx}
\usepackage{cite}
\usepackage{calc}
\usepackage[fleqn]{amsmath}
\usepackage{amssymb}
\usepackage{natbib}
\usepackage{color}

\def\aap{A\&A}
\def\aapr{A\&A Rev.}

\def\apjl{ApJL}

\def\mnras{MNRAS}
\def\aj{AJ}

\def\vesc{v_{\rm esc}}
\def\rcrit{r_{\rm crit}}
\def\rhatcrit{\hat{r}_{\rm crit}}

\usepackage[normalem]{ulem}
\usepackage{soul}

\definecolor{mycolor}{rgb}{1,0.5,0}

\title[Spherical models of star clusters with potential escapers]{Spherical models of star clusters with potential escapers}
\author[Ian Claydon, Mark Gieles, Anna Lisa Varri, Douglas C. Heggie, Alice Zocchi]{Ian Claydon$^{1}$\thanks{E-mail:
i.claydon@surrey.ac.uk (IC); m.gieles@surrey.ac.uk (MG)}, Mark
Gieles$^{1,2,3}$, Anna Lisa Varri$^{4}$, Douglas C. Heggie$^{5}$, Alice Zocchi$^{6}$ \\
$^{1}$Department of Physics, University of Surrey, Guildford, GU2 7XH, UK \\
$^2$Institut de Ci\`{e}ncies del Cosmos (ICCUB), Universitat de Barcelona, Mart\'{i} i Franqu\`{e}s 1, E08028 Barcelona, Spain\\
$^3$ICREA, Pg. Lluis Companys 23, 08010 Barcelona, Spain.\\
$^{4}$Institute for Astronomy, University of Edinburgh, Royal Observatory, Blackford Hill, Edinburgh EH9 3HJ, UK \\
$^{5}$School of Mathematics and Maxwell Institute for Mathematical Sciences, University of Edinburgh, The King's Buildings, Edinburgh EH9 3FD \\
$^{6}$European Space Research and Technology Centre (ESA/ESTEC), Keplerlaan 1, 2201 AZ Noordwijk, The Netherlands}
\begin{document}

\date{}

\pagerange{\pageref{firstpage}--\pageref{lastpage}} \pubyear{2018}

\maketitle

\label{firstpage}

\begin{abstract}
An increasing number of observations of the outer regions of globular clusters (GCs) have shown a flattening of the velocity dispersion profile and an extended surface density profile. Formation scenarios of GCs can lead to different explanations of these peculiarities, therefore the dynamics of stars in the outskirts of GCs are an important tool in tracing back the evolutionary history and formation of star clusters. One possible explanation for these features is that GCs are embedded in dark matter halos. Alternatively, these features are the result of a population of energetically unbound stars that can be spatially trapped within the cluster, known as potential escapers (PEs). We present a prescription for the contribution of these energetically unbound members to a family of self-consistent, distribution function-based models, which, for brevity, we call the Spherical Potential Escapers Stitched ({\sc spes}) models. We show that, when fitting to mock data of bound and unbound stars from an $N$-body model of a tidally-limited star cluster, the {\sc spes} models correctly reproduce the density and velocity dispersion profiles up to 
the Jacobi radius, and they are able to recover the value of the Jacobi radius itself to within 20\%. We also provide a comparison to 
the number density and velocity dispersion profiles of the Galactic cluster 47 Tucanae. Such a case offers a proof of concept that an appropriate modeling of PEs
is essential 
to accurately interpret current and forthcoming \textit{Gaia} data in the outskirts of GCs, and, in turn, to formulate meaningful present-day constraints for 
GC formation scenarios in the early universe.

\end{abstract}


\section{Introduction}
\label{Sect:Intro}

\subsection{Globular clusters as quasi-isothermal systems}

Globular clusters (GCs) are 
ancient stellar systems orbiting around the centre of mass of their host galaxies. Their evolution is the result of two-body relaxation, stellar evolution, binary star evolution and the interaction with the galactic tidal field \citep[e.g.][]{1997A&ARv...8....1M}. Despite the complex interplay of these processes, their present day properties are well captured by relatively simple dynamical models  \citep[e.g.][]{Gunn1979}. As progressively more accurate observational data unveils the complexities of GCs' structural and 
kinematic properties, advances to these simple models have been required to accurately describe them. Understanding the physical processes that generate these complexities may hold the key to understanding the formation process and evolution of GCs.

Dynamical models of GCs are usually of two types. First, evolutionary models, e.g. numerical simulations based on direct $N$-body \citep{Nitadori2012,Wang2015} and Monte Carlo approaches \citep{Freitag2001,Giersz2006}, include many of the complex aspects of GC evolution. These take into account, among other physical ingredients, the collisional nature of the systems, stellar evolution and the perturbations induced by the galactic environment, which provide a realistic description of these systems. However, million-particle $N$-body models tailored to describe the observational properties of individual clusters, although finally achievable \citep{Heggie2014,Wang2016}, still require a significant investment of computational time.

An alternative modeling approach is to use equilibrium models which describe the properties of clusters at a given time in their evolution. An example of these types of models are those defined by a distribution function (DF), describing the density of points in phase space. These models are faster to solve than evolutionary models, and provide a simple but physically justified description of the bulk internal properties of GCs.
We refer to \citet{VHB2019} for a comparison of the performance of several equilibrium models (e.g., DF-based or moments-based) in the interpretation of mock surface brightness, radial velocity and proper motion profiles derived from a reference $N$-body model of the cluster M4 \citep{Heggie2014}.

The most popular class of DF-based models of GCs are the so-called `lowered isothermal' models, which are approximately isothermal in the central regions, but have a finite escape velocity to mimic the effect of the energy truncation induced by the galactic tidal field. Anisotropy in the velocity distribution can be found in GCs as a consequence of their conditions at formation \citep{Vesperini2014,Breen2017}, or as a product of their evolution \citep{Oh1992,Baumgardt2003}, and recent numerical simulations of star clusters showed that anisotropy evolves during the lifetime of GCs, depending also on the initial conditions, including how compact the cluster is \citep{Sollima2015,Tiongco2016}. For these reasons, including the presence of anisotropy in the models \citep[e.g., in the way proposed by][]{Eddington1915,Michie1963} can be important to accurately reproduce evolutionary effects \citep[e.g., see][]{Zocchi2016} and, most crucially, observations (e.g., see Anderson \& van der Marel 2010, Watkins et al. 2015). The effects of mass segregation can be taken into account by incorporating several components in the models \citep{DaCosta1976,Gunn1979}, to describe the dynamics of stars with different masses. The possibility to have radially dependent mass-to-light ratios and anisotropy has proven important in the discussion on intermediate-mass black holes in GCs \citep{Illingworth1977,Zocchi2017,Gieles2018,Zocchi2019} and dark remnants \citep{Sollima2015,Sollima2016,Peuten2017,Zocchi2019}.

Recently, \citet{Gieles2015} developed the {\sc limepy} family of models which are isothermal at low energies and polytropic near the truncation energy. The truncation of the models is controlled by the parameter $g$, allowing for the truncation prescription to vary smoothly between the ones proposed by \citet{Woolley1954}, \citet{King1966} and (non-rotating) \citet{Wilson1975} modes. The {\sc limepy} models include a prescription for radial velocity anisotropy \citep[for a test against $N$-body models, see][]{Zocchi2016}, and the possibility to consider multiple mass components, which accurately reproduce the phase-space distribution of multimass $N$-body models \citep[][]{Peuten2017}. Additionally, the inclusion of differential rotation (e.g., by means of a prescription equivalent to the one adopted by \citealt{Prendergast1970}) is straightforward (Zocchi \& Varri, in prep.), but comes with a higher computational cost because of the loss of spherical symmetry.

\subsection{Old and new observables and their possible dynamical interpretation}
Despite the developments summarised above, there are complexities in the observational data that can not be reproduced by existing models. These include a flattening of the velocity dispersion profile near the Jacobi radius $r_{\rm J}$ (\citealt{Drukier1998}; \citealt{Scarpa2007}; \citealt{Lane2009}), extended haloes (\citealt{Cote2002}; \citealt{Olszewski2009}; \citealt{CarballoBello2012}; \citealt{Kuzma2016,Kuzma2018}), and high velocity stars (\citealt{Meylan1991}; \citealt{Lutzgendorf2012}; \citealt{Kamann2014}).

Traditional expectations of Newtonian dynamics in the outskirts of GCs would suggest a decreasing velocity dispersion profile with increasing radius. Earlier and more recent empirical evidence suggests that, in some Galactic clusters, the velocity dispersion may be elevated and the surface density may be raised compared to this expectation \citep{Drukier1998,Lane2011,Carballo-Bello2018} which has led some to propose the inclusion of additional physics beyond Newtonian predictions. These explanations include modified theories of gravity \citep{Hernandez2011}, where once a star passes below a threshold value in acceleration (and also the clusters orbital velocity around the galaxy is below the same threshold) the star can enter a modified dynamics regime \citep{Milgrom1983}, which can lead to a flat velocity dispersion at large distances from the cluster centre. Alternatively, some formation scenarios suggest that GCs could form within their own dark matter mini-haloes, similar to dwarf galaxies \citep{Peebles1984,Mashchenko2005,Trenti2015}. If still present, this would elevate the velocity dispersion \citep[e.g., see][]{Ibata2013,Penarrubia2017}. 

Other formation theories, where GCs  formed in gas-rich discs and major mergers of galaxies do not require the presence of a dark matter halo \citep{Kravtsov2005}. In this scenario, the peculiarities could be explained by the GCs still being in the debris of a disrupted dwarf galaxy after a merger \citep{Carballo-Bello2018}. 

Finally, the tidal field of the host galaxy introduces a spatial condition for escape in addition to a critical energy for escape. This leads to the presence of a population of stars with an energy above the critical energy for escape but still spatially bound to the cluster, and within $r_{\rm J}$. The effects of these so-called PEs were first investigated with $N$-body models after it was found that the dissolution time of simulations with large particles numbers were shorter than in scaled up simulations with smaller particle numbers \citep{Fukushige&Heggie2000}. The expectation was that the dissolution time of clusters scales linearly with the half-mass relaxation time, $t_{\rm rh}$, because this is the time-scale for stars to be scattered above the critical energy for escape, $E_{\rm crit}$. However, a dependency of $t_{\rm rh}^{3/4}$ for the dissolution time  was found in $N$-body models of star clusters evolving in steady tidal fields \citep{Baumgardt2001}. This deviation from a linear dependence on $t_{\rm rh}$ can be understood from the additional timescale of the spatial criteria for escape through one of the Lagrangian points \citep{Baumgardt2001}, which is dependent on the cluster mass. This means PEs can persist within the cluster for a long time before escaping, where in ideal circumstances some can even remain inside a cluster indefinitely \citep{Henon1969}.

PEs dominate in the outer regions of clusters, and around a radius of half of the Jacobi radius ($0.5\,r_{\rm J}$), roughly 50\% of the stars are PEs \citep{Kupper2010}. \citet[][from here on C17]{Claydon2017} showed that the total amount of PEs depends on the assumed shape of the stellar mass function and the galactic potential. Due to the fact that their energy is larger than that of the other stars in the cluster, PEs contribute to increasing the velocity dispersion profile and to extending the surface density profile beyond the extent of bound stars. 
The effects of PEs can only account for the behaviour of the surface density profiles inside of $r_{\rm J}$. However, by including these effects in an equilibrium model such as the one described in this study, the resulting estimate of the Jacobi radius $r_{\rm J}$ is much more accurate than the one based on simpler, `lowered isothermal' models \citep[see][]{deBoer2019}.
This makes them a possible explanation for the peculiarities in observational data \citep{Kupper2010} and the amount of deviation from the Newtonian expectation for the velocity dispersion. In addition, their spatial properties can be used to infer properties of the dark halo of their host galaxy.

Therefore, observationally determining if these peculiarities are due to PEs or dark matter can constrain the formation scenario and evolutionary processes that shape GC dynamics. ESA's {\it Gaia} mission is providing a revolutionary set of data, with positions and proper motions for a billion stars in the Galaxy. This includes the previously unprobed population of stars in the outskirts of GCs. It is therefore paramount to understand the effect that PEs can have on the observations, and to propose a model that accounts for their behaviour.
\begin{figure}
	\centering
	\includegraphics[width=\columnwidth]{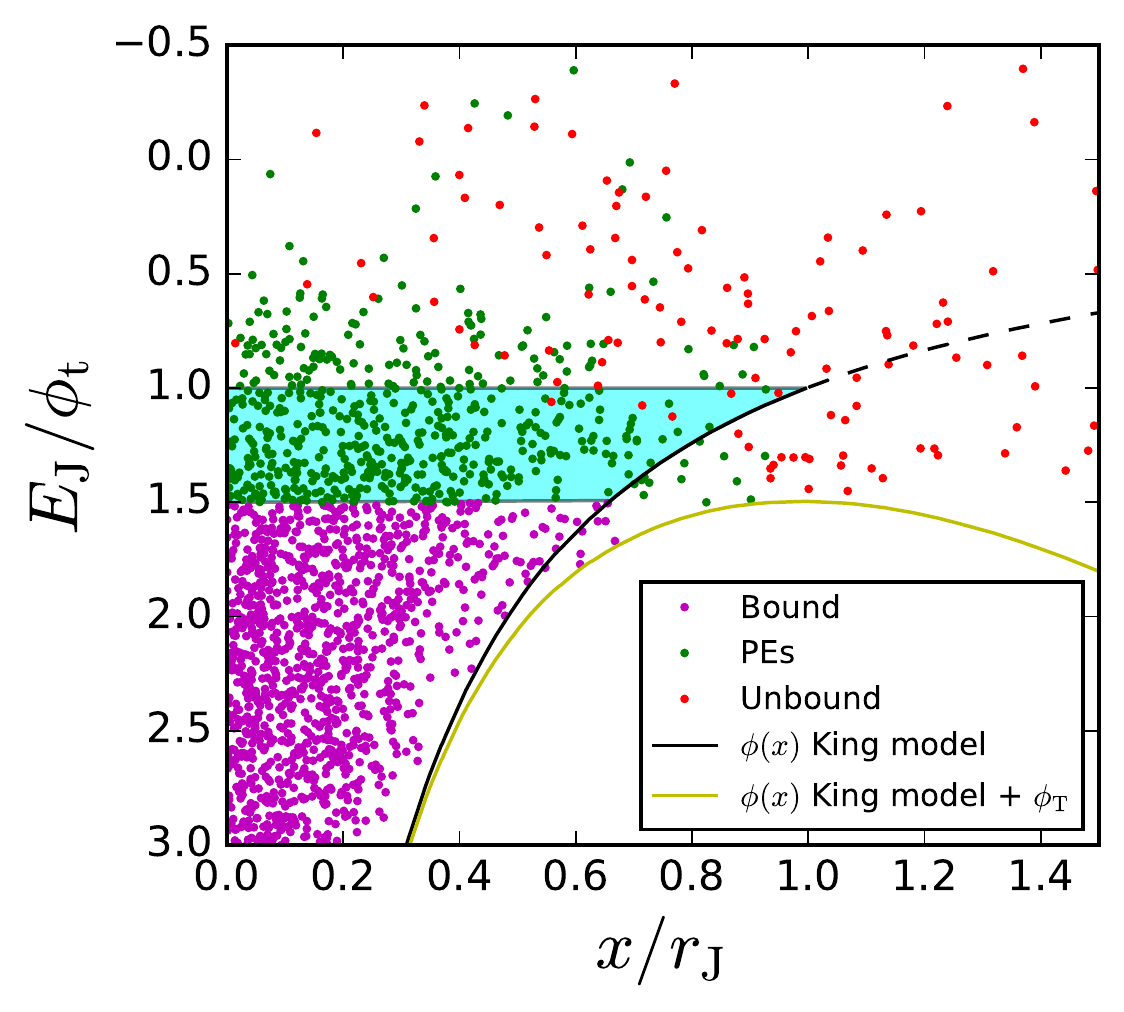}
	\caption{Jacobi energy ($E_{\rm J}$) normalised to the critical energy, as a function of the position of stars on the $x$-axis normalised to $r_{\rm J}$ from an $N$-body models from C17 (ss3, as described in Section 2). Magenta points are bound stars ($E_{\rm J} < E_{\rm crit}$), green points are PEs ($E_{\rm J} > E_{\rm crit}$ and $r<r_{\rm J}$) and red points are unbound stars ($r>r_{\rm J}$). The potential of the King model (black) and the King model plus tides (yellow) are shown. The shaded cyan region shows the range of PEs which are included in the King model fit, but would no longer be included if the effects of the galactic potential are introduced. }
	\label{fig:intro1}
\end{figure}
\subsection{Adding unbound stars to bound models}
\citet*{Daniel2017} developed a family of DF-based models of GCs that include the effects of PEs, which are described in terms of approximate integrals of motion, as inspired by a family of periodic orbits of the circular Hill problem proposed by \citet{Henon1969}. Unfortunately, such an approach does not allow to easily formulate a simple analytical expression of the DF, which, in turn, makes the derivation of a fully self-consistent solution of the relevant Poisson equation quite cumbersome. The family of models presented in this paper partly addresses these two limitations, although at the cost of introducing substantial simplifications in the phase space description of the PEs.

Given the importance of PEs in $N$-body models, one may wonder why traditional models (without PEs), such as King's model, offer a satisfactory representation of the observational properties of many GCs. This is partially because most King model fits are done to data that does not extend all the way to $r_{\rm J}$ \citep{Trager1995}. Another reason is that the truncation energy of such models does not necessarily correspond to the critical energy of the systems they describe. This allows the existing models to account for the presence of some PEs inside of the model, albeit with incorrect underlying physics. This is because these models are isolated, and the effect of the tides is mimicked by `lowering' the energy by a truncation energy $\phi_{\rm t}=-GM_{\rm c}/r_{\rm t}$, with $G$ the gravitational constant, $M_{\rm c}$ the cluster mass and $r_{\rm t}$ the truncation radius. This energy
is larger than what the critical energy for escape would be if the effects of a galactic tidal potential are included (defined here as $E_{\rm crit}$). 
For a cluster on a circular orbit in a reference frame corotating with the orbit, a star has a Jacobi energy  of
\begin{equation}
E_{\rm J} = \frac{v^2}{2} + \phi_{\rm c} + \frac{\Omega^2}{2}(z^2 - 3x^2) \ ,
\label{eq:ej}
\end{equation}
where the terms on the right hand side are the kinetic energy, the potential energy due to the cluster, a contribution from the galactic tidal potential, and the centrifugal force \citep{Fukushige&Heggie2000}. The critical energy of the system \citep{Heggie2003} is
\begin{equation}
E_{\rm crit}=-\frac{3GM_{\rm c}}{2r_{\rm J}} \,.
\end{equation}
This means that an isolated model when fit to data from $N$-body simulations of GCs (orbiting in a corotating reference frame around a time-independent galactic potential) will describe some PEs with $-1.5\lesssim E_{\rm J}/\phi_{\rm t}\lesssim -1$ as bound members of the system.

We illustrate this by comparing a King model to the energy of stars in a tidally limited $N$-body model. We fit a King model to a snapshot from an $N$-body model from C17, on a circular orbit around a singular isothermal galactic potential, and compare the potential from the model at any radius, $\phi(r)$, to the $E_{\rm J}$ of each star. We define bound stars as stars with $E_{\rm J}<E_{\rm crit}$, PEs as stars with $r<r_{\rm J}$ and $E_{\rm J} > E_{\rm crit}$ and unbound stars  as stars with $r>r_{\rm J}$. Figure~\ref{fig:intro1} plots $E_{\rm J}$ normalised to $\phi_{\rm t}$, against the $x$-axis position normalised to $r_{\rm J}$, with the bound stars shown in magenta, PEs in green and unbound in red. By fixing the mass to be the correct value and $r_{\rm t}=r_{\rm J}$ we fit on the concentration parameter of the King model and plot the potential (black line) which denotes the minimum energy a star can have at that radius. The potential beyond $r_{\rm t}$ is approximated as a point mass with  $\phi(x)=-GM/x$ (dashed line). We also plot $\phi(x) + \phi_{\rm T}$ where we have added the tidal and centrifugal contribution: $\phi_{\rm T}= -1.5\Omega^2x^2$ (yellow line, see equation~\ref{eq:ej}). PEs with $E_{\rm crit} < E_{\rm J} < \phi_{\rm t}$ (in the shaded cyan region) are included in the King model ($\sim73\%$ of PEs), and the model could include more PEs by increasing $r_{\rm t}$. Therefore the model is able to reproduce the bulk properties of the data but does not have the correct underlying physics to describe the dynamics.

The goal of this study is to develop a convenient, spherically-symmetric family of models that include an approximate description of the phase space contribution of PEs, which are within the Jacobi radius of GCs. Such a family is defined by a distribution function formulated as a simple analytical expression, which can agilely allow to derive a self-consistent solution of the corresponding Poisson equation. In Section 2 we describe the models and explore their properties. In Sections 3 and 4 we compare the models to $N$-body simulations and observational data, respectively. In Section 5 we discuss the strengths and limitations of the models and delineate how the approximations taken may be improved upon in future versions of the models.

\section{The {\sc spes} family of models}

\subsection{Distribution function}
\begin{figure}
	\centering
	\includegraphics[width=\columnwidth]{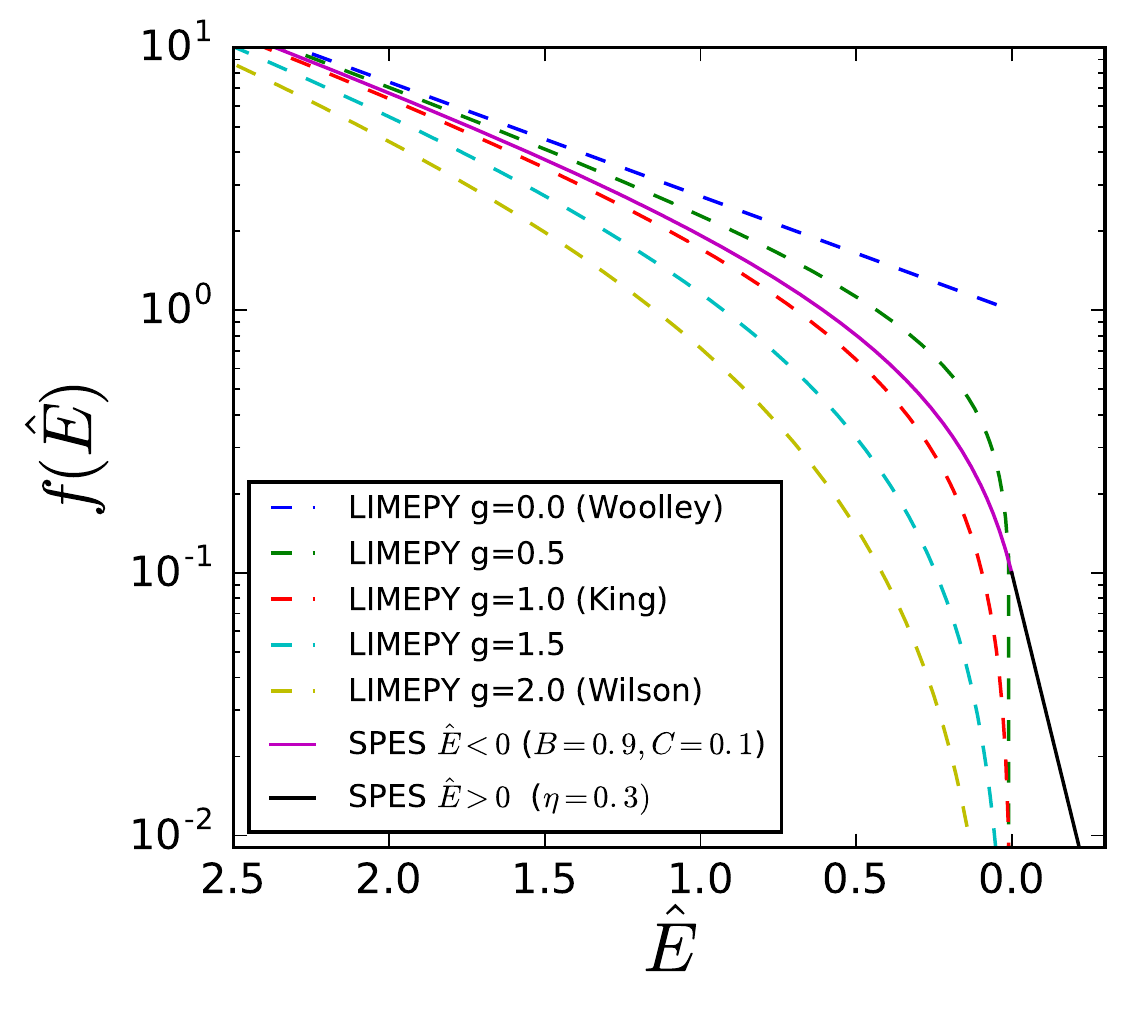}
	\caption{Distribution function as a function of energy for {\sc limepy} models (dashed lines) with $g=0$, 0.5, 1, 1.5, and 2, and for the bound (purple solid line) and unbound (black solid line) parts of the {\sc spes} model (equation~\ref{eqn:df}) with $B = 0.9$ and $\eta = 0.3$. All models have $\hat{\phi}_0=9$.}
	\label{fig:df}
\end{figure}

One of the earliest lowered isothermal models used a purely isothermal DF, $f(\hat{E}) = A\exp(\hat{E})$ for $\hat{E}>0$ and $f(\hat{E})=0$ elsewhere \citep{Woolley1954}, where $\hat{E}= -(E-\phi_{\rm t})/s^2$, $E=0.5v^2+\phi(r)$ is the specific energy and stars with negative $\hat{E}$ are assumed to instantaneously escape. This family of models is characterized by two physical scales: the normalisation
of the DF (i.e., the free constant $A$), which sets the mass of the system, and a velocity scale $s$. This function is discontinuous at $\hat{E}=0$ (i.e. $E=\phi_{\rm t}$). 

To ensure a vanishing phase space density and a continuous DF at $\hat{E}=0$, a constant can be subtracted from the exponential, such that: $f(\hat{E})=A\left[\exp(\hat{E})-1\right]$ \citep{King1966}\footnote{The \citet{King1966} model is in fact an approximate steady state solution of the Fokker-Planck equation, when considering two-body relaxation and escape. A concise explanation of the physical justification of the King model can be found in \citet{King2008}.}. The resulting density profiles in projection match the surface brightness profiles of GCs exceptionally well \citep{Trager1995}, which is why this approach has been the foundation of many further developments.

The models proposed by \citet{Wilson1975}, as taken in the non-rotating and isotropic limit, include an additional energy term, $f(\hat{E}) =A\left[\exp(\hat{E})-1 - \hat{E}\right]$. This makes the derivative of $f(\hat{E})$ continuous at $\hat{E}=0$, and leads to a more extended density distribution which has been shown to better fit observed density profiles of some GCs \citep[see the discussion in][]{McLaughlin2005}.

\citet{Davoust1977} and \citet{Hunter1977} showed that Woolley, King and Wilson models are special members of a family of models with different orders of truncation of the isothermal DF; more recently, an updated formulation of the DF by \citet{gomezleyton2014} also allows to construct solutions in between these models. The DF is defined as $f(\hat{E}) = A \, \exp(\hat{E}) \, \gamma(g,\hat{E})/\Gamma(g)$, where $\gamma(a,x)$ and $\Gamma(x)$ are the lower incomplete gamma function and the  gamma function, respectively. When considering $g=0, 1$ and $2$, the Woolley, King and Wilson models are obtained, respectively. 
This was further developed in the {\sc limepy} family of models by \citet{Gieles2015}, who added radial orbit anisotropy and multiple mass components and provided a {\sc python} implementation\footnote{{\sc limepy} is available from https:/github.com/mgieles/limepy}.

The construction of DF-based models which include a contribution from a population of PEs may be conducted by adopting the following rationale. First, we rely on the simplifying assumptions of equilibrium and spherical symmetry. Second, concerning the representation of the phase space behaviour of the bound population, we choose to preserve some consistency with the class of lowered isothermal models described above, as they offer an  empirically satisfactory description of the dynamics of the central regions of many Galactic globular clusters.
We recognise that the assumption of dynamical equilibrium introduces a significant degree of idealisation in our description of the problem. Nonetheless, we emphasise that as shown in \citet{Baumgardt2001}, the distribution of PEs is relatively constant with time, with the predominant evolution being the width of the energy distribution above $E_{\rm crit}$. Therefore, we argue that a static model should be able to match the instantaneous behaviour of PEs at a given time, provided that a parameter setting the energy distribution width is included. 

The rationale adopted above allows us to provide a zeroth-order description of the effects induced by the presence of a population of unbound stars. To achieve this we develop a spherically symmetric distribution function that only depends on energy, which has the additional advantage of preserving a certain mathematical simplicity and rapidity of numerical calculation. In the future, we intend to address the limitation of ignoring the anisotropy in the velocity dispersion and the deviations from spherical symmetry introduced by the effects of a galactic potential (as discussed also in Section 5), by considering the constructions of DF-based models
which take into account the non-spherical nature of the external tidal field and other dynamical ingredients. 

Unfortunately, the {\sc limepy} models are not suited to add an unbound population. Figure~\ref{fig:df} shows the DF as a function of the energy, for several isotropic {\sc limepy} models and different values of $g$ (dashed lines). For values of $g>0$, the DF vanishes at $\hat{E}=0$. This means a discontinuity would be introduced when including the effects of PEs. To solve this, in our definition of the DF we introduce the constants $B$ and $C$, which control the value of the DF and its derivative at $\hat{E}=0$. This is an approach similar to the one used for Woolley ($B=C=0$), King ($B=1$, $C=0$) and Wilson models ($B=C=1$), but in this case we leave the values of these parameters free, to have a non-zero density at $\hat{E}=0$ for $B<1$. 

This approach allows us to `stitch' to the DF for bound stars the one for PEs, with the `stitching' taking place at $\hat{E}=0$. The choice of the functional form 
adopted for the DF of PEs 
is motivated by numerical results  from extensive direct $N$-body investigations. 
It has been shown that the  evolution of the number of stars, $N(\hat{E})$, for $\hat{E}<0$ is well described by a modified-Bessel function, when the effects of dynamical friction are not included \citep{Baumgardt2001}. However it is not possible to derive an analytic expression for $f(\hat{E})$ from this $N(\hat{E})$. Therefore we approximate it by an isothermal model in the regime $\hat{E}<0$. A more rigorous approach may be taken in future versions of the models, but the approximation of an exponential DF for the PEs is adequate for the fits shown in Section 3 and 4.

The DF of the {\sc spes} (Spherical Potential Escaper Stitched) family of models is
\begin{equation}
		\displaystyle
        f(\hat{E}) = A\times\begin{cases}
                \exp(\hat{E})-B - C\hat{E}, & \text{$\hat{E} \geq 0$}\\
 \displaystyle (1-B) \exp\left(\frac{\hat{E}}{\eta^2}\right), & \text{$\hat{E}<0$} \ ,
        \end{cases}
        \label{eqn:df}
\end{equation}
where $\eta^2 = s_{\rm pe}^2 / s^2$, where $s_{\rm pe}$ is the 1D velocity dispersion of the PEs. 

Once the stars become energetically unbound, their escape time $t_{\rm e} \propto \hat{E}^{-2}$  \citep{Fukushige&Heggie2000}. This means that $t_{\rm e}(\hat{E}=0) = \infty$ and that the stars only slightly above the critical energy have very large escape times and the effects of the escape process are negligible. This suggests that the DF should therefore be continuous across $\hat{E}=0$ and also continuous in the derivative to ensure that the behaviour of stars slightly above and slightly below $\hat{E}=0$ is similar. Enforcing continuity to further derivatives would over-constrain the model as the DF only has terms to second order. Additionally the simple isothermal model assumed for the DF for $\hat{E}<0$ is a reasonable approximation for the zeroth and first derivatives but it is likely inaccurate for further derivatives. If we then demand smoothness, we find
\begin{equation}
	C = 1-\frac{1-B}{\eta^2}.
        \label{eqn:two}
\end{equation}
A representative example of the behaviour of the DF for the case of a model with $B=0.9$, $\eta=0.3$ is illustrated in Fig.~\ref{fig:df}.

In addition to the usual degree of freedom which controls the central concentration ($\hat{\phi}_0$, e.g., see King 1966), the PE-specific parameters of the model are $\eta$ and $B$, which define $C$ via equation~(\ref{eqn:two}). Moreover, as in the case of the conventional `lowered isothermal' models, two physical scales can be set by means of the free constants $s$ and $A$ (e.g., the velocity and mass scale for the system).
Acceptable values for the $\eta$ parameter are between 0 and 1, to ensure that the value of the velocity dispersion at the tidal radius of the model assumes values between 0 and (approximately) the central value of the velocity dispersion. Also, for the other parameter, we impose $0\leq B\leq1$ so that the density at $\hat{E}=0$ can vary between a non-zero maximum (for $B=0$) to zero (for $B=1$). It follows that $-\infty < C \leq1$, and because the derivative of the DF at $\hat{E}=0$ is proportional to $1-C$, its range is $0\leq f^\prime(0)\leq\infty$.  We note that the (non-rotating, isotropic) Wilson model is found for $B=1$, regardless of the value of $\eta$, and the King model is recovered for $B=1$ and $C = 0$. 

As the model is no longer `truncated' in the same way as previous models, we refer to the critical radius $r_{\rm crit}$ as the radius where the specific potential reaches the critical value $\phi=\phi_{\rm t}$ which is therefore the maximum radius of bound stars. This parameter is comparable to $r_{\rm t}$ of {\sc limepy} models and $r_{\rm J}$ of data and $N$-body models.

By analysing the outcome of numerical simulations, \citet{Claydon2017} showed that during the lifetime of the cluster the distribution of PEs within the Jacobi radius maintains the same shape, with only the width of the energy distribution above $E_{\rm crit}$ changing significantly. This suggests that the {\sc spes} models could be able to reproduce the instantaneous properties of PEs thanks to the parameter $\eta$, which is related to the width of the energy distribution. However, care must be taken when using this model as initial conditions for evolutionary modeling. By introducing this unbound contribution from the DF the model will no longer be in virial equilibrium, and the model is unstable unless the effects of a specific galactic potential are included.  The only way to include a galactic potential is by including an impermeable boundary at $r_{\rm crit}$ and the model will then be in equilibrium  when considering the total kinetic energy, $K$, the total potential energy, $W$, and a pressure term\footnote{The boundary contributes to the radial component of the pressure tensor, see Section~\ref{Sect:Disc}}: $p_{\rm t}V =s_{\rm pe}^2 \rho(r_{\rm crit})(4/3)\pi r_{\rm crit}^3$, such that the condition for virial equilibrium is $2K - W - 3p_{\rm t}V = 0$ \citep{LyndenBellWood1968}. 

\begin{figure*}
	\centering
	\includegraphics[width=0.85\textwidth]{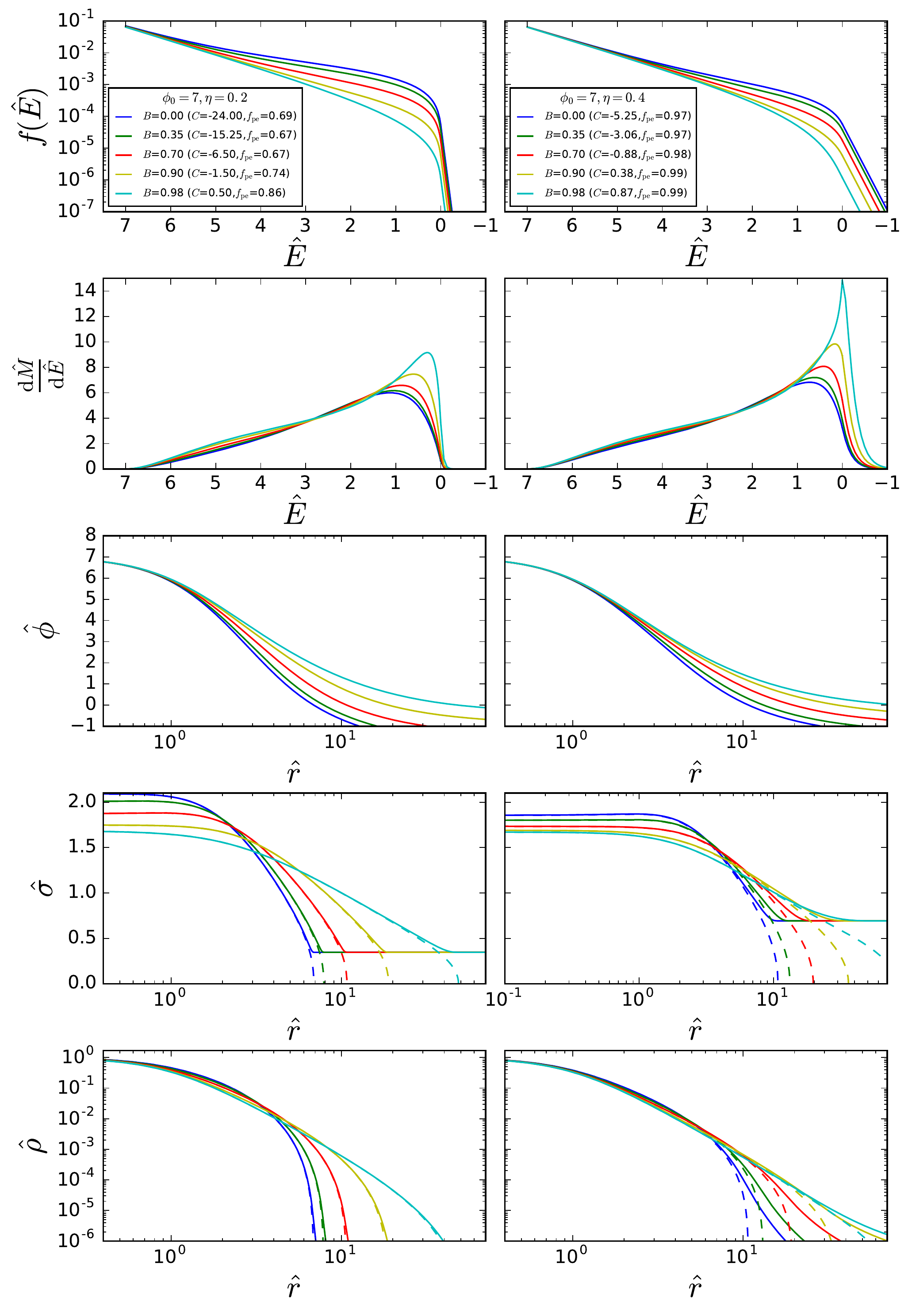}
    \caption{{\sc spes} model properties while varying the input parameters. All panels have $\hat{\phi}_0$=7, left-hand panels have $\eta$=0.2, right hand panels $\eta$=0.4 and all panels vary $B$ from 0.1 to 0.98, which in turn varies $C$ from -23.75 to 0.5 and -5.19 to 0.8 for $\eta$=0.2 and 0.4 respectively. Solid lines are the complete profiles, dashed lines are just the bound contribution to the DF. The quantities displayed are the distribution function $f(\hat{E})$ (first row) and the differential energy distribution ${\rm d}\hat{M}/{\rm d}\hat{E}$ (i.e., the amount of mass per unit energy; second row) profiles against $\hat{E}$, and the potential $\hat{\phi}$ (third row), the velocity dispersion $\hat{\sigma}$ (fourth row) and the density $\hat{\rho}$ (fifth row) profiles against $\hat{r}$.}
	\label{fig:diag2}
\end{figure*}

\subsection{Properties of the models}

To compute the models, we define the dimensionless quantities $\hat{\phi} = (\phi_{\rm t}-\phi(r))/s^2$, $\hat{r} = r/r_{\rm s}$, and $\hat{\rho}=\rho/\rho_0$ \citep[see also][]{King1966,Gieles2015}, where  $\rho_0$ is the central density and $r_{\rm s}^2 = 9s^2/(4 \pi G \rho_0)$ is the (square of the) scale radius, or King radius. 
The radius at which $\hat{\phi}= 0$ is $\rcrit$. The Poisson equation for the dimensionless potential $\hat{\phi}$ can be written as:
\begin{equation}
	\frac{1}{\hat{r}^2}\frac{\rm d}{\rm{d}\it\hat{r}}\left(\hat{r}^2\frac{\rm d \it \hat{\phi}}{\rm d \it \hat{r}}\right) = -9\hat{\rho},
\end{equation}
which can be solved by assuming the following boundary conditions at $\hat{r}=0$: $\hat{\phi}=\hat{\phi}_0$, ${\rm d}\hat{\phi}/{\rm d}\hat{r}=0$, where $\hat{\phi}_0$ is a positive constant defining the dimensionless parameter which sets the central concentration of a model (this parameter is called $W_0$ in \citealt{King1966}). The density and pressure as a function of $\hat{\phi}$ can be found from 
\begin{equation}
\rho =\int f(E) {\rm d}^3v =(2\pi s^2)^{3/2}A \mathcal{I}^{\rho} 
\end{equation}
and
\begin{equation}
\rho\sigma^2 =\int f(E)v^2 {\rm d}^3v = 3(2\pi s^2)^{3/2}s^2A\mathcal{I}^{\rho\sigma^2}. 
\end{equation}
Here the integration over all velocities is split in a regime $0\le v\leq\vesc$ for the bound stars and $\vesc<v<\infty$ for the PEs. Here $\vesc$ is the escape velocity required to move from bound to potential escaper regime, i.e. $\vesc = \sqrt{2(\phi_{\rm t}-\phi(r))}$. We introduced dimensionless density and pressure integrals which are given by
\begin{equation}
		\mathcal{I}^{\rho} =E_{\gamma}\left(\frac{3}{2},\hat{\phi}\right) - \frac{B\hat{\phi}^{3/2}}{\Gamma(5/2)} 
	 - \frac{C\hat{\phi}^{5/2}}{\Gamma(7/2)}  + (1-B)\eta^3E_{\Gamma}\left(\frac{3}{2},\frac{\hat{\phi}}{\eta^2}\right) 
\end{equation}

\begin{multline}
	\mathcal{I}^{\rho\sigma^2}= E_{\gamma}\left(\frac{5}{2},\hat{\phi}\right) - \frac{B\hat{\phi}^{5/2}}{\Gamma(7/2)} - \frac{C\hat{\phi}^{7/2}}{\Gamma(9/2)}+ (1-B)\eta^5 E_{\Gamma}\left(\frac{5}{2},\frac{\hat{\phi}}{\eta^2}\right).
\end{multline}
Here we used the previously introduced function  $E_{\gamma}(a,x)=\exp(x)\gamma(a,x)/\Gamma(a)$ \citep[see][]{Gieles2015}  and introduce $E_{\Gamma}(a,x) = \exp(x)\Gamma(a,x)/\Gamma(a)$, where $\Gamma(a,x)$ is the upper incomplete Gamma function. The normalised density is found by dividing by the central density $\mathcal{I}^{\rho}_0 = \mathcal{I}^{\rho}(\hat{\phi}_0)$, 
\begin{equation}
\hat{\rho} = \frac{\mathcal{I}^{\rho}}{\mathcal{I}^{\rho}_0} ,
\end{equation}
and the velocity dispersion is obtained as
\begin{equation}
	\hat{\sigma} = \sqrt{3\frac{\mathcal{I}^{\rho\sigma^2}}{\mathcal{I}^{\rho}}}.
\end{equation}
where $\hat{\sigma}=\sigma/s^2$. If $\eta$ is very small, this gives rise to a very large argument of the exponential  resulting in numerical problems. Therefore, for values of $x>700$ we replace $E_{\Gamma}(a,x)$ with its limiting behaviour for large $x$: $x^{a-1}/\Gamma(a)$. 

We can also obtain surface density profiles $\Sigma(\hat{R})$ and line-of-sight velocity dispersion profiles, $\sigma_{\rm LOS}(\hat{R})$, where $\hat{r}^2=\hat{R}^2+\hat{Z}^2$ and $\hat{Z}$ is along the line-of-sight: 
\begin{equation}
        \hat{\Sigma}(\hat{R}) = 2\int_{0}^{D\hat{r}_{\rm crit}}\hat{\rho} d\hat{Z}
\end{equation}
and
\begin{equation}
        \hat{\sigma}_{\rm LOS}^2(\hat{R}) = \frac{2}{\hat{\Sigma}(\hat{R})}\int_{0}^{D\hat{r}_{\rm crit}}\hat{\rho}(\hat{r})\frac{\hat{\sigma}^2(\hat{r})}{3} d\hat{Z}.
\end{equation}
where we limit the integral along $\hat{Z}$ to a multiple of $\hat{r}_{\rm crit}$ by defining the fitting parameter $D$, which we discuss in $S3.1$.

\subsubsection{Limits and mass}

In the regime, $\hat{r} \rightarrow \hat{r}_{\rm crit}$, $\hat{\phi} \rightarrow 0$, the density and velocity integrals are
\begin{equation}
\lim_{\hat{\phi} \rightarrow 0}\mathcal{I}^{\rho} = (1-B)\eta^3\left(1+\frac{\hat{\phi}}{\eta^2} +\frac{\hat{\phi}^2}{2\eta^4} \right) + \mathcal{O}(\hat{\phi}^{5/2})
\end{equation}
and
\begin{equation}
\lim_{\hat{\phi} \rightarrow 0}\mathcal{I}^{\rho\sigma^2} = (1-B)\eta^5\left(1+\frac{\hat{\phi}}{\eta^2} +\frac{\hat{\phi}^2}{2\eta^4}+\frac{\hat{\phi}^3}{6\eta^6} \right) + \mathcal{O}(\hat{\phi}^{7/2}).
 	\label{eqn:limits}
\end{equation}

We can also solve the model beyond $\hat{r}_{\rm crit}$, where $\hat{\phi}\leq 0$, to compare the model to data including unbound stars beyond $r_{\rm J}$. In this regime there is no contribution from $f(\hat{E}>0)$ and the integration boundary $\vesc=0$, therefore the density and velocity dispersion simplify to:
\begin{equation}
	\mathcal{I}^{\rho}(\hat{r}>\hat{r}_{\rm crit}) = (1-B)\eta^3\exp(\hat{\phi}/\eta^2) ,
\end{equation}
and
\begin{equation}
	\mathcal{I}^{\rho\sigma^2} (\hat{r}>\hat{r}_{\rm crit})=\eta^2 \mathcal{I}^{\rho}(\hat{r}>\hat{r}_{\rm crit}),
\end{equation}
such that the mean-square velocity is a constant: $\hat{\sigma}^2(\hat{r}>\hat{r}_{\rm crit}) = 3\eta^2$.

The mass of the model inside $\rhatcrit$ is calculated as $\hat{M}_{\rm c}= \int_0^{\rhatcrit} 4 \pi \hat{r}^2\hat{\rho}{\rm d}\hat{r}$. The contribution to the mass from the unbound part of the DF is $\hat{M}_{\rm pe} = \int_0^{\rhatcrit} 4 \pi \hat{r}^2 \hat{\rho}_{\rm pe} {\rm d}\hat{r}$, with $\hat{\rho}_{\rm pe} = (1-B)\eta^3E_{\Gamma}(3/2,\hat{\phi}/\eta^2)/\mathcal{I}^{\rho}_{0}$; the bound contribution to the mass is therefore $\hat{M}_{\rm b} = \hat{M}_{\rm c} - \hat{M}_{\rm pe}$ and the fraction of mass in PEs is $\mathcal{F}_{\rm pe} = \hat{M}_{\rm pe}/\hat{M}_{\rm c}$. 
The {\sc spes} models are available in the {\sc limepy} package from https://github.com/mgieles/limepy\footnote{After installing {\sc limepy} the {\sc spes} models can be imported in {\sc python} as: {\texttt{from limepy import spes}}}.

\subsection{Exploring the parameter space}
\begin{figure}
	\centering
	\includegraphics[width=\columnwidth]{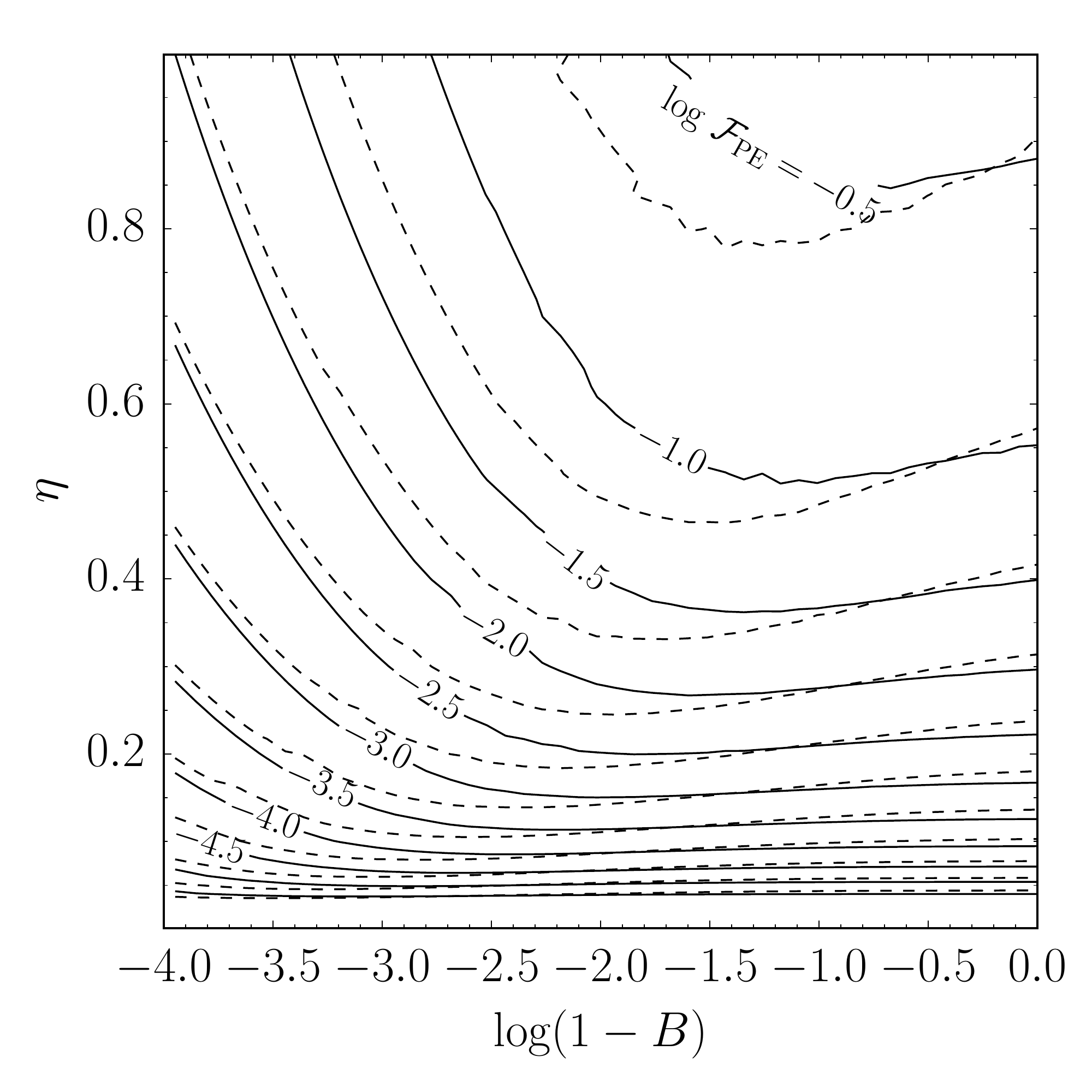}
	\caption{Fraction of PEs for different combinations of $B$ and $\eta$ and $\hat{\phi}_0=5$ (full lines) and $\hat{\phi}_0 = 6$ (dashed lines).}
	\label{fig:fpe}
\end{figure}
\begin{figure}
	\centering
	\includegraphics[width=\columnwidth]{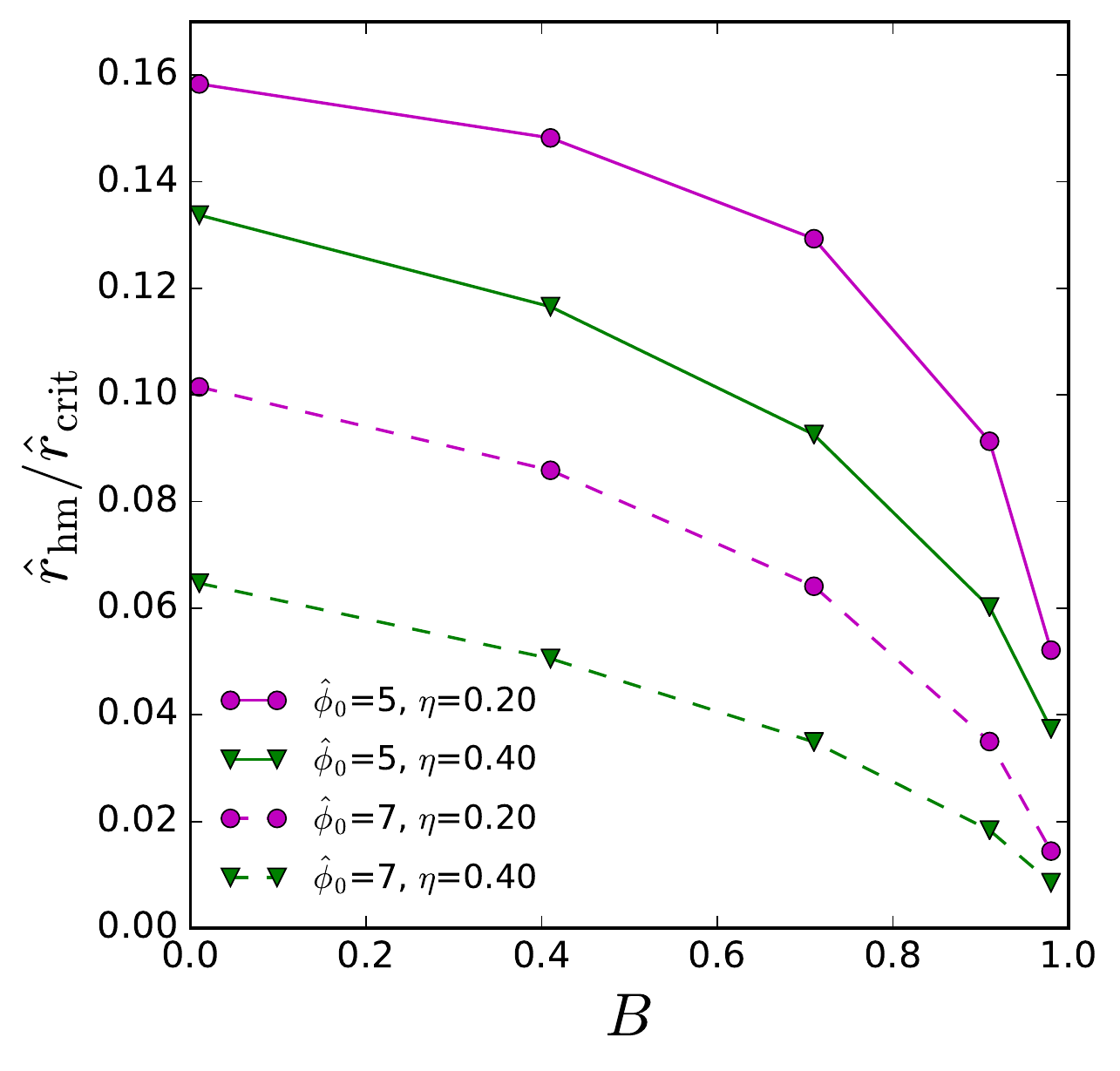}
	\caption{Values of the ratio between the half-mass and truncation radius $\hat{r}_{\rm hm}/\hat{r}_{\rm crit}$ against $B$, for $\hat{\phi}_0=7$ (dashed lines), $\hat{\phi}_0=5$ (solid lines) and $\eta=0.2$ (magenta) and 0.4 (green).}
	\label{fig:diag3}
\end{figure}
We solve Poisson's equation by splitting it into two first order ordinary differential equations, and by using a Runge-Kutta integrator with an adaptive step-size, {dopri5} \citep{Hairer1993}. We consider different values of the parameters $\hat{\phi}_0$, $\eta$ and $B$ to investigate the behaviour of the model. All figures and analysis in this section are presented in the dimensionless model units. 

The parameter $B$ controls the phase space density at $\hat{E}=0$, which, in turn, controls the truncation of the model. When increasing $B$ (while keeping the other parameters fixed), $\rhatcrit$ and the mass increase (Fig.~\ref{fig:diag2}). The parameter $\eta$ sets the ratio of the value of the velocity dispersion at $\rhatcrit$ to the velocity scale $s$, which for high values of $\hat{\phi}_0$ approaches the (one-dimensional) central velocity dispersion. However, because $C$ also affects the truncation of the model and is a function of $B$ and $\eta$, changing $\eta$ will also vary $\rhatcrit$ and changing $B$ can also change $\hat{M}_{\rm pe}$. 

For a fixed $B$, the PE fraction increases with increasing $\eta$. This is because increasing $\eta$ makes the DF wider, thus increasing the mass in PEs. 
The dependence of the PE fraction on $B$ is not as trivial. Because the phase space density at the critical energy is proportional to $1-B$, we expect the PE fraction to correlate with $1-B$. This is true for $B\simeq1$, but for smaller $B$ the two quantities anticorrelate (at fixed $\eta$), while for $\eta\simeq0$, the fraction of PEs is approximately independent of $B$. The PE fraction depends also on $C$ in the DF, which is determined by the demand for continuity and smoothness. To illustrate how the PE fraction depends on the model parameters, we show in Fig.~\ref{fig:fpe}  the fraction of PEs in models with $\hat{\phi}_0=5$ (solid lines) and  $\hat{\phi}_0=6$ (dashed lines) and different combinations of $B$ and $\eta$. 

Figure~\ref{fig:diag3} shows the values of the ratio between the half-mass and truncation radius $\hat{r}_{\rm hm}/\hat{r}_{\rm crit}$ against $B$, for different values of $\eta$ and $\hat{\phi}_0$. By inspecting the figure, it appears that this quantity is a monotonically decreasing function of $B$, with an additional dependence on $\eta$ which is more significant for low values of $B$.

\begin{table*}
\begin{center}
    \caption[Properties of the best-fit {\sc limepy} and {\sc SPES} models to $N$-body data]{Properties of the best-fit models. For each model, indicated in the first column, we provide: the central potential $\hat{\phi}_0$, the model parameters $g$, $\eta$ and $B$, the cluster mass $M_{\rm c}$, the half-mass radius $r_{\rm hm}$, the Jacobi radius $r_{\rm J}$, the mass of bound stars $M_{\rm b}$, the mass of PEs $M_{\rm pe}$, and the ratio of potential escaper mass to total cluster mass $\mathcal{F}_{\rm pe}$. Rows are the simulations ($N$-body), the best-fit {\sc limepy} models, and the best-fit {\sc spes} models to 3D data within $r_{\rm J}$, within $2r_{\rm J}$ and projected on the $xy$, $yz$ and $xz$ axes for each snapshot ss1, ss2, ss3 and ss4 from the C17 simulation.}
\scalebox{1}{\begin{tabular}{|l|l|c|c|c|c|c|c|c|c|c|c|c|}
 \hline
 Model & snapshot & $\hat{\phi}_0$ &  $g$ & $\eta$ & $B$ &  $M_{\rm c}$ &$r_{\rm hm}$ &$\rcrit$ & $M_{\rm b}$ & $M_{\rm pe}$ & $\mathcal{F}_{\rm pe}$\\ \hline\hline
$N$-body & ss1          & - &  -& - & - & 0.800 & 0.623& 6.12& 0.753& 0.048& 0.060  \\ \hline
{\sc limepy} & ss1      & 7.70 &  1.24 & - & - & 0.789 & 0.606 & 7.38 & - &  - & -\\
{\sc spes} & ss1        & 7.24 &-& 0.201 & 0.974 & 0.796 & 0.592 & 6.08 & 0.801 & 0.006 & 0.008  \\
{\sc limepy} & ss1.2rj  & 7.40 &  1.55 & - & - & 0.777 & 0.570 & 10.39 & - &  - & -\\
{\sc spes} & ss1.2rj     & 7.68 &-& 0.267 & 0.933 & 0.801 & 2.057 & 5.02 & 0.779 & 0.015 & 0.019  \\
{\sc spes} &  xy ss1.2rj & 7.71 & - &0.290 & 0.911 & 0.790 & 0.675 & 5.06 & 0.771 & 0.019 & 0.024\\
{\sc spes} &  xz ss1.2rj & 7.73 &- & 0.293 & 0.910 & 0.794 & 0.617 & 4.62 & 0.774 & 0.020 & 0.026\\
{\sc spes} &  yz ss1.2rj & 7.80 &- & 0.273 & 0.916 & 0.793 & 0.809 & 6.13 & 0.775 & 0.015 & 0.019 \\\hline\hline

$N$-body & ss2           & -  & -& -    & - & 0.601 & 0.794& 5.56& 0.540& 0.061& 0.102 \\ \hline
{\sc limepy} & ss2       & 10.70 & 1.44 & - & - & 0.597 & 0.771 & 8.05 &  - & - & -\\
{\sc spes} & ss2         & 11.26 &- & 0.226 & 0.982 & 0.600 & 0.781 & 5.63 & 0.589 & 0.013 & 0.022  \\
{\sc limepy} & ss2.2rj   & 11.30 & 1.73 &  - & - & 0.594 & 0.760 & 10.98 &  - & - & -\\
{\sc spes} & ss2.2rj     & 10.24 &- & 0.251 & 0.959 & 0.606 & 2.118 & 4.97 & 0.578 & 0.018 & 0.030 \\
{\sc spes} &  xy ss2.2rj & 10.16 &- &0.262 & 0.955 & 0.592 & 0.851 & 5.18 & 0.571 & 0.021 & 0.035  \\
{\sc spes} &  xz ss2.2rj & 10.35 &- & 0.263 & 0.956 & 0.595 & 0.914 & 5.56 & 0.574 & 0.021 & 0.035 \\
{\sc spes} &  yz ss2.2rj & 10.50 &- & 0.257 & 0.955 & 0.598 & 0.882 & 5.31 & 0.580 & 0.018 & 0.030 \\\hline\hline

$N$-body & ss3           & - & - & - & - & 0.400 & 0.808& 4.86& 0.346& 0.054& 0.136 \\\hline
{\sc limepy} & ss3       & 12.10 & 1.52 & - & - & 0.399 & 0.790 & 8.40  &  - & - & -\\
{\sc spes} & ss3         & 11.68 &-& 0.276 & 0.967 & 0.399 & 0.795 & 5.09 & 0.388 & 0.015 & 0.037 \\
{\sc limepy} & ss3.2rj   & 11.60 & 1.67 & - & - & 0.400 & 0.784 & 10.23 &  - & - & -\\
{\sc spes} &  ss3.2rj    & 10.72 & -&0.290 & 0.927 & 0.407 & 1.992 & 4.50 & 0.384 & 0.015 & 0.038 \\
{\sc spes} &  xy ss3.2rj & 10.56 & - &0.289 & 0.940 & 0.395 & 0.818 & 4.60 & 0.379 & 0.017 & 0.043 \\
{\sc spes} &  xz ss3.2rj & 10.80 &- & 0.277 & 0.947 & 0.396 & 0.827 & 4.77 & 0.382 & 0.014 & 0.036\\
{\sc spes} &  yz ss3.2rj & 11.22 &- &0.286 & 0.942 & 0.401 & 0.813 & 4.63 & 0.388 & 0.014 & 0.035\\\hline\hline

$N$-body & ss4 & -  & -& - & - & 0.201 & 0.746& 3.86& 0.168& 0.034& 0.166 \\ \hline
{\sc limepy} & ss4 & 10.90 & 1.47 & - & - & 0.201 & 0.710 & 7.26  & - & - & - \\
{\sc spes} & ss4 & 10.78 &-& 0.304 & 0.942 & 0.201 & 0.708 & 4.17 & 0.190 & 0.010 & 0.051  \\
{\sc limepy} & ss4.2rj & 10.90 & 1.74& - & - & 0.204 & 0.805 & 10.67  & - & - & - \\ 
{\sc spes} & ss4.2rj & 10.28 &-& 0.290 & 0.949 & 0.208 & 1.786 & 4.01 & 0.192 & 0.010 & 0.049 \\
{\sc spes} &  xy ss4.2rj & 10.26 &- & 0.286 & 0.958 & 0.200 & 0.726 & 4.38 & 0.190 & 0.010 & 0.051 \\
{\sc spes} &  xz ss4.2rj & 10.54 &- & 0.300 & 0.953 & 0.203 & 0.804 & 4.66 & 0.192 & 0.011 & 0.056 \\
{\sc spes} &  yz ss4.2rj & 10.45 &- & 0.294 & 0.929 & 0.204 & 0.784 & 4.26 & 0.196 & 0.009 & 0.043 \\\hline\hline

    \end{tabular}}
\end{center}
\end{table*}

\section{Fitting to $N$-body simulations}

To test the performance of the {\sc spes} models in describing globular cluster properties we fit the {\sc spes} models to snapshots from $N$-body simulations of tidally limited star clusters. For comparison, we also fit all $N$-body models with {\sc limepy} models. We consider the simulations presented in C17 which describe systems with $N=16384$ equal-mass stars. The model clusters are evolved on a circular orbit around the centre of mass of their host galaxy, which is spherically symmetric and characterized by a power-law mass distribution $M_{\rm g}(<R_{\rm g})\propto R_{\rm g}^{\lambda}$, where $R_{\rm g}$ is the galactocentric distance. We consider the cases where $\lambda=1$, which correspond to singular isothermal sphere. The data from the simulations are analysied in a corotating reference frame, where the $x$-axis is along the direction linking the centre of the cluster and the centre of the galaxy and the $y$-axis is in the direction of the tangential component of the orbital angular velocity vector \citep{Heggie2003}. The simulations were run using \textsc{nbody6tt}, which allows a functional input for the galactic potential (\citealt{Aarseth2003}; \citealt{Nitadori2012}; \citealt{Renaud2015}). The data from the simulations are in H\'{e}non units \citep{Henon1971}, where $G=1$, the initial mass of the clusters $M_{\rm c0}=1$ and total energy of the cluster $\mathcal{E}_{0} = -1/4$ . The analysis presented in this section is computed in these units. 

\subsection{Fitting technique}

We calculate the velocity dispersion and density profiles by binning the data of four snapshots, corresponding to the moments during the lifetime of the simulations when the remaining mass is 0.8, 0.6, 0.4 and 0.2 of the initial mass. We compute the profiles by considering bins with equal numbers of stars and by taking into account all stars within the Jacobi radius, $r_{\rm J}$, (ss1, ss2, ss3 and ss4) and for all stars within  $2r_{\rm J}$ (ss1.2rj, ss2.2rj, ss3.2rj and ss4.2rj).

We fit the models to these density and velocity dispersion profiles by using a Markov Chain Monte Carlo technique, {\sc emcee} \citep{Foreman2013}, to explore the parameter space of the DF-based models. The best-fit values of the parameters are obtained by minimizing the associated $\chi$-squared: 
\begin{equation}
        \chi^2 = \sum_{i=1}^{n} \frac{(O_i-\mathcal{M}_i)^2}{\epsilon_i^2},
\end{equation} 
where $O_i$ are the data values, $\epsilon_i$ are the errors on the data values, in this case the standard error from calculating the $\sigma$ and $\rho$ profiles from the $N$-body data, and $\mathcal{M}_i$ are the model values at the same radial position as the $n$ data values. We calculate this both for the density and for the velocity dispersion. 

The parameters are $\hat{\phi}_0$, $B$ and $\eta$ and two scale values to convert the model units $\hat{M}$ to H\'{e}non units $M_{\rm c}$, $M_{\rm scale} = M_{\rm c}/\hat{M}$, and $r_{\rm scale} = r_{\rm hm} / \hat{r}_{\rm hm}$ and we stop the model at the radius where the potential $\hat{\phi}=0$ which we call the critical radius $r_{\rm crit}$. When fitting to data beyond $r_{\rm J}$, as the model is infinite this will elevate the surface density profile when projecting the model. Therefore we require a stopping radius further out than $r_{\rm crit}$ and we define $\hat{r}_{\rm stop} = D \hat{r}_{\rm critt}$ and redefine $r_{\rm scale} = r_{\rm lb}/\hat{r}_{\rm stop}$ where $r_{\rm lb}$ is the radius of the last bin of data. By fitting on $D$ we can then allow $\hat{r}_{\rm crit}$ to be any value less than $\hat{r}_{\rm stop}$.For each parameter, we determine the best-fit value as the median of the correspondent marginalised posterior probability distribution, and $1\sigma$ errors as the 16 and 84 per cent percentiles.
\begin{figure}
	\centering
	\includegraphics[width=\columnwidth]{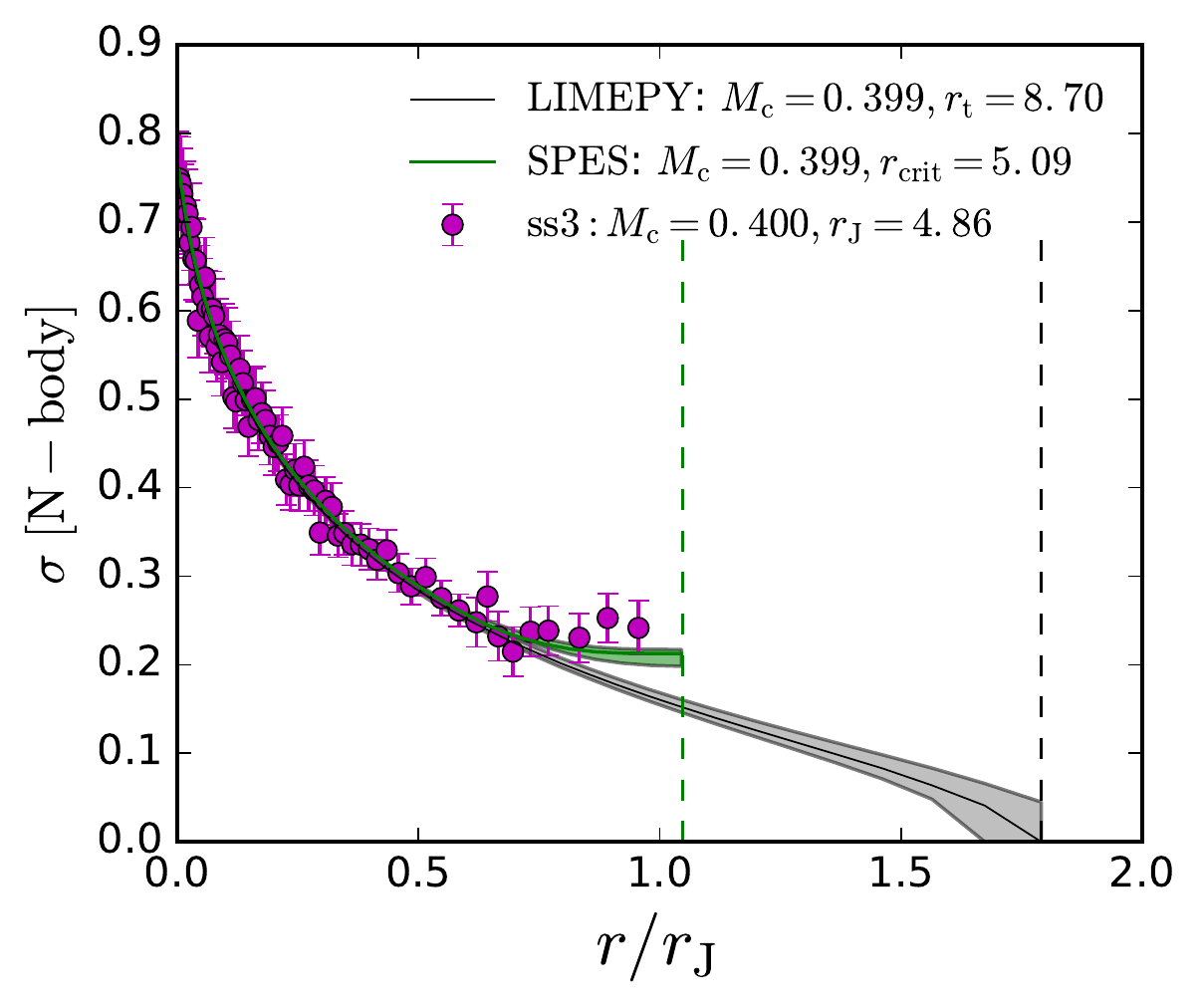}
	\caption{Velocity dispersion profile from ss3 (magenta points) against $r$, normalised to $r_{\rm J}$. The black and green lines represent the best-fit {\sc limepy} and {\sc spes} model, respectively. The shaded grey and green region represent models that occupy a $1\sigma$ region around the maximum likelihood, as identified by {\sc emcee}. The vertical green and black dashed line indicate the best-fit truncation radii of {\sc spes} and {\sc limepy} models, respectively. The {\sc spes} model is able to closely match $r_{\rm J}$ whereas {\sc limepy}  overestimates it.}
	\label{fig:BC_limepy}
\end{figure}
\subsection{Fitting to 3D profiles}

In this section we describe the results we obtained when fitting the models to the snapshots by considering 3D profiles. We conduct this test to assess the ability of the {\sc limepy} model and of the {\sc spes} model to reproduce the properties of the snapshots when having all the possible information, i.e. 6D data (all dimensions of the configuration space and velocity space) for the considered stars.
\subsubsection{Stars within $r_{\rm J}$}

\begin{figure*}
	\centering
	\includegraphics[width=0.91\textwidth]{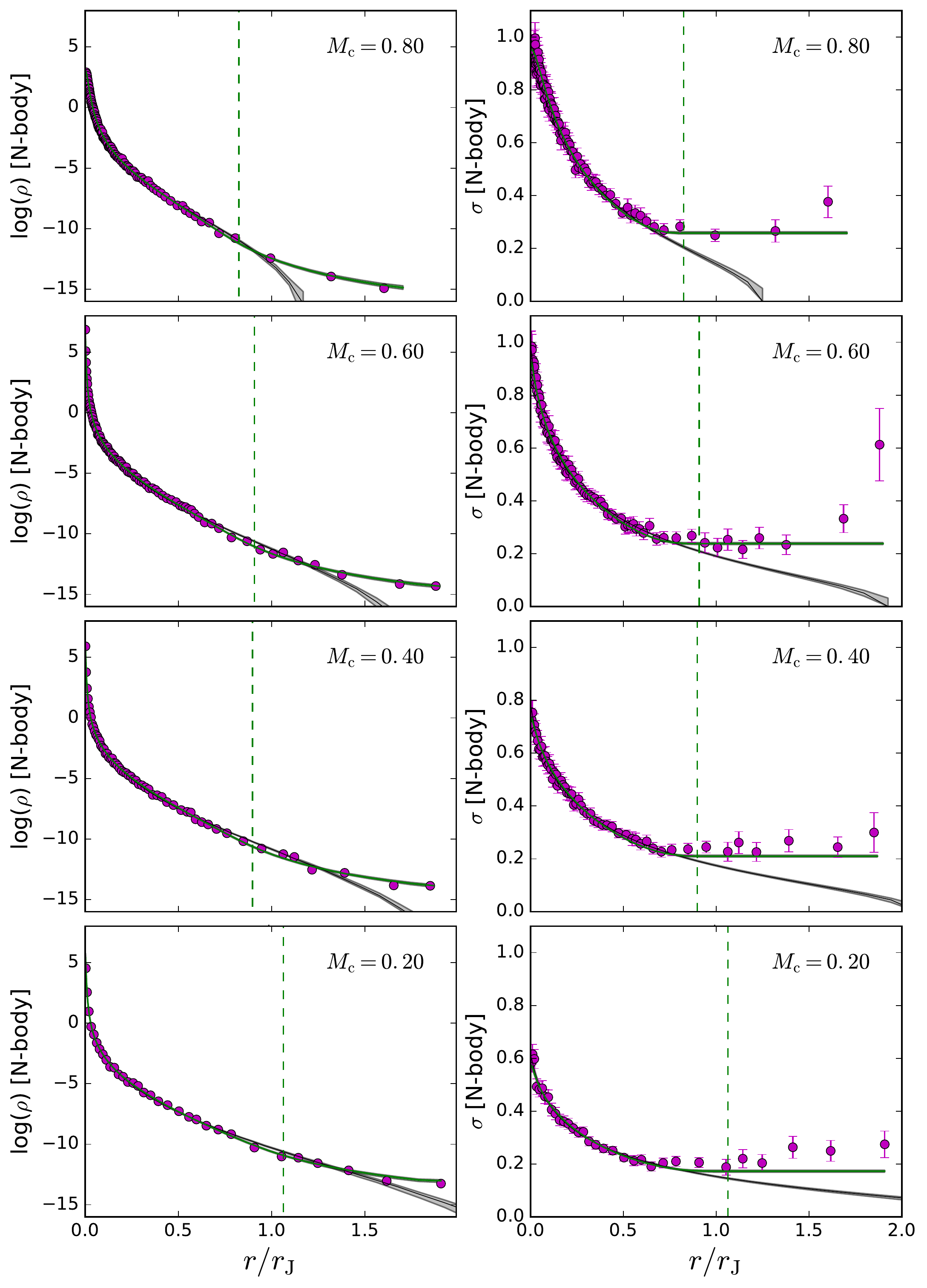}
	\caption{Three-dimensional density $\rho$ (left-hand panels) and velocity dispersion $\sigma$ (right-hand panels) profiles as a function of radius in units of $r_{\rm J}$ constructed from binned data (magenta points) from the four snapshots from the $\lambda=1$ C17 simulation. The best-fit {\sc spes} and {\sc limepy} models are displayed in the same way as Fig.~\ref{fig:BC_limepy}. Vertical dashed green lines are the $\rcrit$ of the best-fit model, showing that even without prior knowledge of $r_{\rm J}$ the {\sc spes} models can reproduce it within $\sim 20\%$.}
	\label{fig:BC2}
\end{figure*}

We first limit our analysis to stars within $r_{\rm J}$ of the $N$-body model. The first part of Table 1 shows the values of the best-fit parameters and the properties of the snapshots for comparison.

 The {\sc limepy} model fits the data well for $r\lesssim0.7r_{\rm J}$ and it closely reproduces $M_{\rm c}$ and $r_{\rm hm}$, but it cannot account for the behaviour of the velocity dispersion profiles at radii towards and beyond $r_{\rm J}$. Moreover, the resulting best-fit value for $\rcrit$ overestimates $r_{\rm J}$ by a factor of $\sim 1.2-1.8$: this is a common issue when fitting these models to this kind of data, as mentioned in Section~\ref{Sect:Intro}. The {\sc spes} model reproduces the innermost part of the density and velocity dispersion profiles as well as the {\sc limepy} model, but, in addition, it is also able to account for the flattening near $r_{\rm J}$, and to reproduce the correct radial extension of the data. To provide an immediate comparison of {\sc spes} models to {\sc limepy} models, we show an example of the results obtained with this fitting procedure. Figure 6 shows the velocity dispersion profile of the snapshot ss3 represented as a function of the radius, normalised to $r_{\rm J}$. The best-fit {\sc limepy} model is shown as a black line, and the grey shaded area represents models that occupy a $1\sigma$ region around the maximum likelihood. The best-fit {\sc spes} model is shown in green, with the green shaded region again denoting models within a $1\sigma$ region around the maximum likelihood. 
 
\subsubsection{Stars within $2r_{\rm J}$}
We fit the models to all the stars contained within $2r_{\rm J}$ from the cluster centre in the same snapshots (ss1.2rj, ss2.2rj, ss3.2rj and ss4.2rj). This test is useful to understand whether the {\sc spes} models are still able to reproduce $r_{\rm J}$ when fit to data which includes stars beyond $r_{\rm J}$, even though the model does not include the underlying physical behaviour of spatially unbound stars. The second and third rows of each part of Table 1 show the best-fit values of the parameters of {\sc limepy} and {\sc spes} models compared to this data. The density and velocity dispersion profiles for each snapshot (magenta points) and the best-fit {\sc limepy} (grey region, black line) and {\sc spes} (green region, green line) models are shown in Fig.~\ref{fig:BC2}.

The {\sc limepy} models accurately recover the $M_{\rm c}$ and $r_{\rm hm}$, however they overestimate $r_{\rm J}$ even more (factor of  $\sim 1.5-2.3)$ than when fit to data within $r_{\rm J}$. Moreover, {\sc limepy} models are unable to match the velocity dispersion and density profiles beyond $r_{\rm J}$.

The {\sc spes} models perform equally well as the {\sc limepy} models in reproducing the quantities $M_{\rm c}$ and $r_{\rm hm}$ of the snapshots. However {\sc spes} models are able to provide a better fit to the density profile and velocity dispersion profiles even beyond $r_{\rm J}$. The {\sc spes} model is not able to account for an increase in the outermost 2 or 3 bins, which are due to the motion of the stars within the tidal tails.
On average, the {\sc spes} model is also able to reproduce $r_{\rm J}$ (dashed green lines), although it underestimates it $\sim20\%$ initially and becomes more accurate for more evolved clusters.

\begin{figure}
	\centering
	\includegraphics[width=\columnwidth]{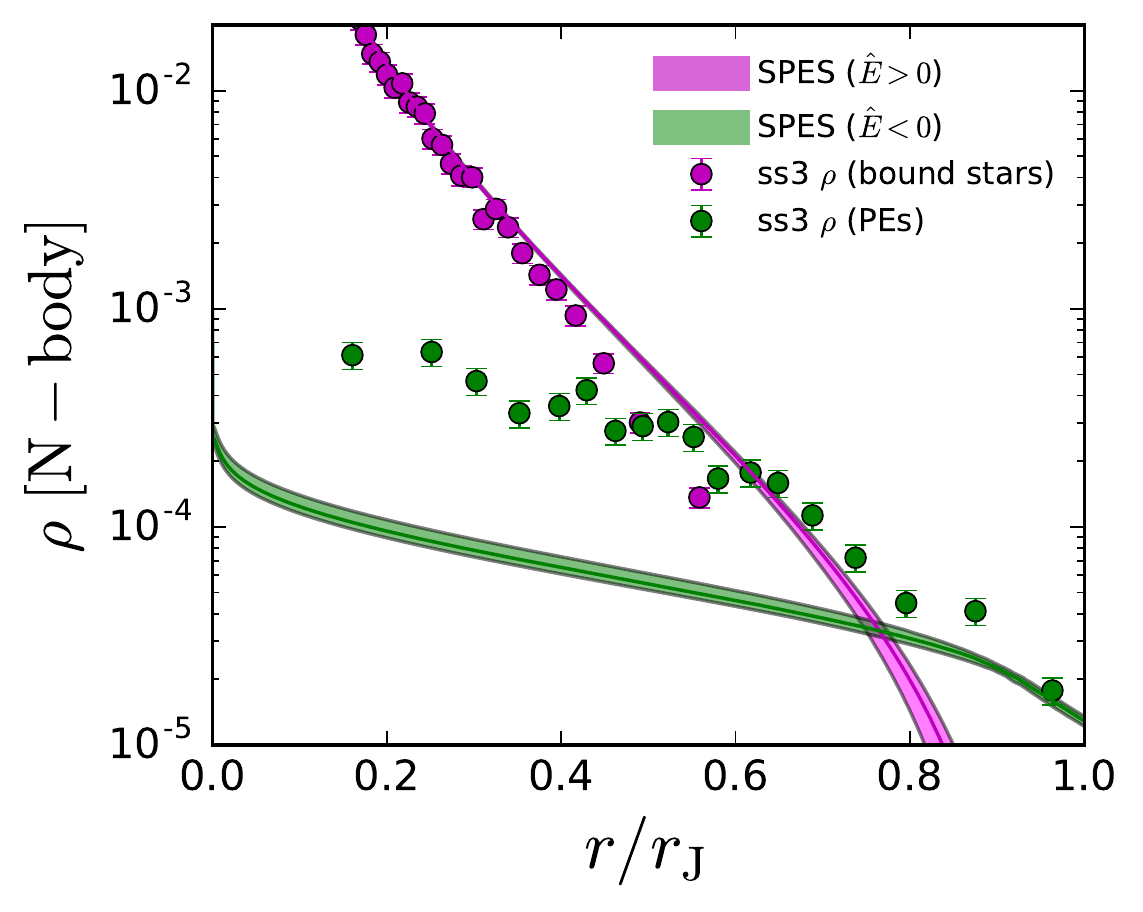}
	\caption{Density profiles for the bound stars (magenta points) and for the PEs (green points) from the snapshot ss3. The bound stars and PEs density profiles of the best-fit {\sc spes} model are shown with the magenta and green lines, respectively. The shaded areas represent models that occupy a $1\sigma$ region around the maximum likelihood.}
	\label{fig:BC3}
\end{figure}

\subsubsection{Fraction of PEs}
\label{Sect:fractionPE}

The {\sc spes} model reproduces the velocity dispersion profile and density profile of the considered snapshots more accurately than models which do not include the contribution of PEs. However, the {\sc spes} model underestimates the fraction of mass in PEs within $\rcrit$, $\mathcal{F}_{\rm pe}$. Table 1 shows that when the {\sc spes} model is fit to the ss.2rj snapshots, it consistently finds a $\mathcal{F}_{\rm pe}$ that is approximately three times lower than the actual value (Table 1; displayed in the $N$-body rows for each snapshot). By separating the density profile into the contribution from bound stars ($\rho_{\rm b}$) and PEs ($\rho_{\rm pe}$) for both the best-fit models and the data, we can see that a large fraction of PEs are actually accounted for by $\rho_{\rm b}$ (Fig.~\ref{fig:BC3}), even when fit to data truncated at $r_{\rm J}$ (ss3), therefore under-predicting $\mathcal{F}_{\rm pe}$ by describing many PEs as bound stars.

This underprediction of $\mathcal{F}_{\rm pe}$ can be attributed to two limitations of the current implementation of the models. First, the approximate expression we assumed for the DF of the PEs does produce a density profile which is, by design, consistent with the density profile of the PEs resulting from direct N-body simulations. Therefore, such a choice may not offer an ideal representation of 
the behaviour of PEs that are only slightly above the critical energy. Second, 
the {\sc spes} models still have a $\phi_{\rm t}$ that is larger than $E_{\rm crit}$ of the stars in the snapshots, as discussed in Section~\ref{Sect:Intro}. Therefore, if $\rcrit= r_{\rm J}$ then PEs with $-3GM_{\rm c}/2r_{\rm J} \lesssim E_{\rm J} \lesssim -GM_{\rm c}/r_{\rm J}$ will be represented by the bound part of the model, and consequently $\mathcal{F}_{\rm pe}$ will be underestimated, even though the total mass is correct. This means that, if the models were to correctly reproduce $\mathcal{F}_{\rm pe}$ without including the galactic potential, it would overestimate $r_{\rm J}$. We conclude that an approximate upward correction of $\mathcal{F}_{\rm pe}$ of a factor of three should be applied to {\sc spes} result to get an estimate of the actual $\mathcal{F}_{\rm pe}$.



\begin{figure}
	\centering
	\includegraphics[width=0.99\columnwidth]{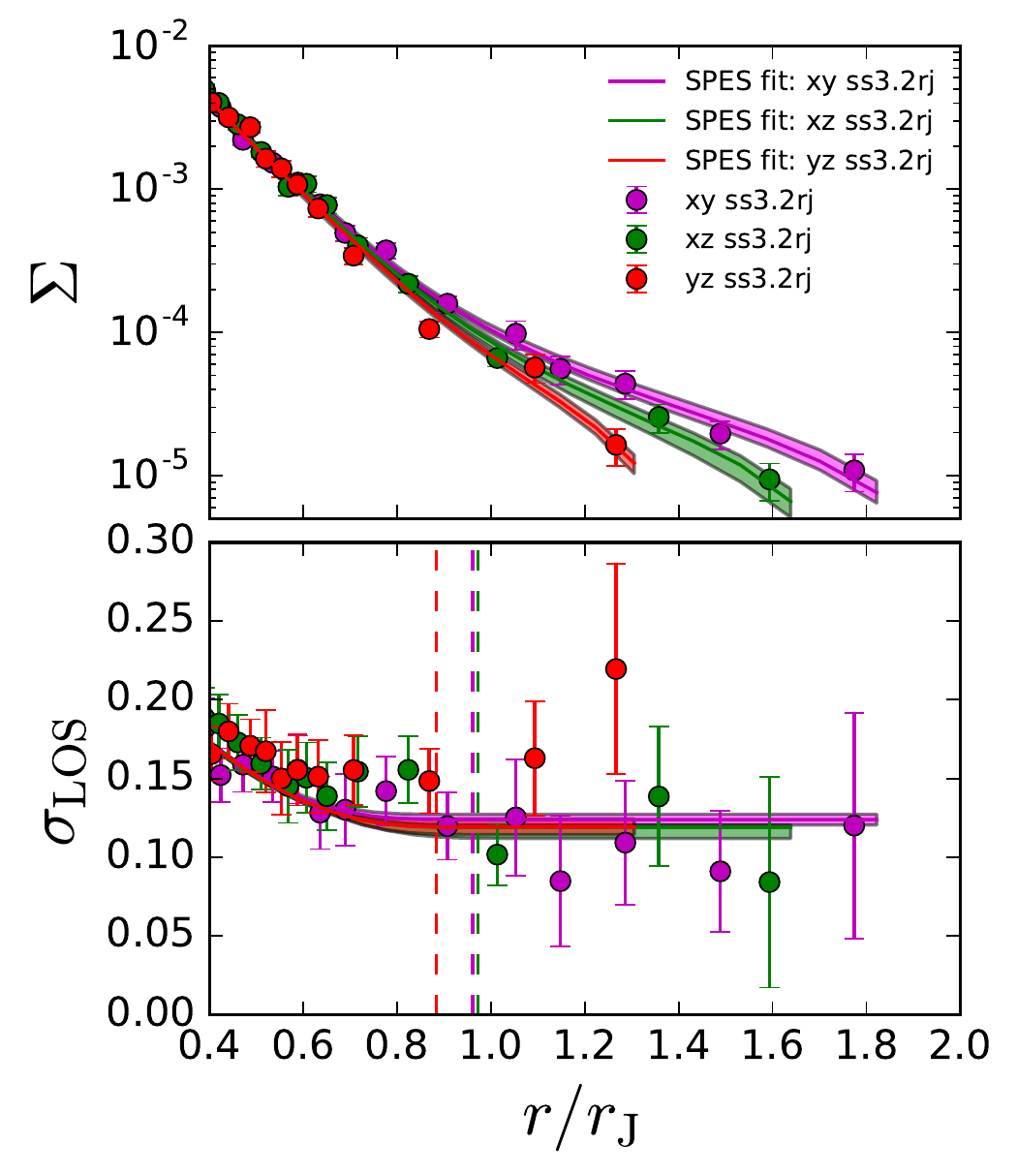}
	\caption{Surface density profiles (top) and projected velocity dispersion profiles (bottom) for the snapshot ss3, projected onto the $(x,y)$ (magenta points), $(x,z)$ (green points) and $(y,z)$ (red points) planes. The best-fit {\sc spes} models are shown as solid lines, and the shaded regions correspond to models within $1\sigma$ of the maximum likelihood; the dashed vertical lines denote the $\rcrit$ of the best-fit models for each projection axis. The colors of the models match the respective data.}
	\label{fig:proj1}
\end{figure}
\begin{figure*}
	\centering
	\includegraphics[width=0.9\textwidth]{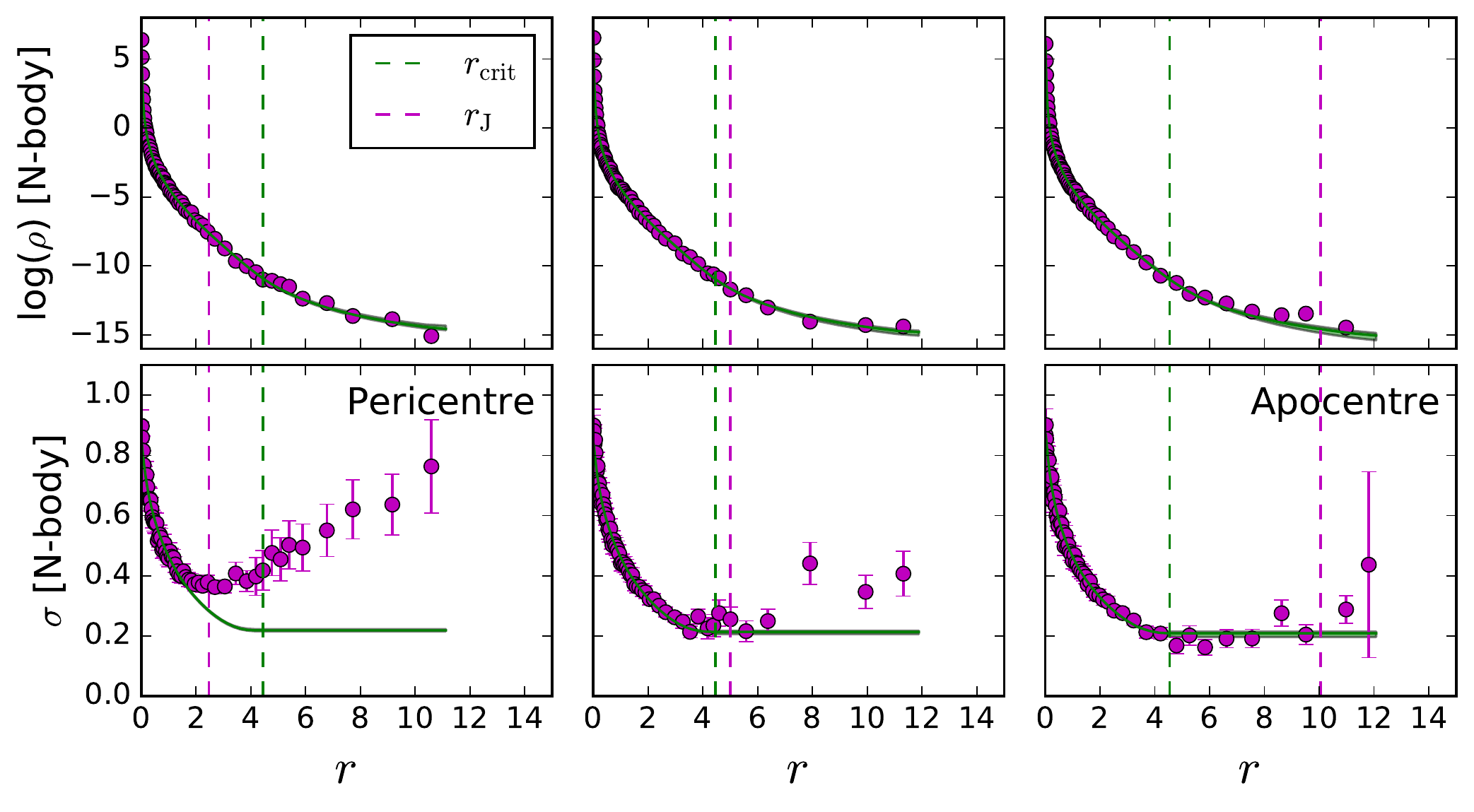}
	\caption{Surface density profiles (top) and projected velocity dispersion profiles (bottom) for three snapshots along one orbit of a simulation of a cluster on an eccentric orbit: pericentre (left-hand panel), apocentre (right-hand panel) and between the two (central panel). The best-fit {\sc spes} models are shown as solid lines, and the shaded regions correspond to models within $1\sigma$ of the maximum likelihood; the dashed vertical lines denote the $\rcrit$ of the best-fit models (green) and $r_{\rm J}$ of the $N$-body data (magenta). 
	}
	\label{fig:eccen}
\end{figure*}

\subsection{Fitting to projected profiles}

In order to test the impact of projection effects on the ability of the models to reproduce the properties of the profiles derived from the $N$-body simulations, we calculate the projections of both the SPES models and the $N$-body data along the line of sight to generate surface density profiles $\Sigma(R)$ and line-of-sight velocity dispersion profiles, $\sigma_{\rm LOS}(R)$, where $r^2 = R^2 + Z^2$ and $Z$ is along the line-of-sight. We consider different directions for the line of sight, and, in particular, we consider the principal axes of our corotating reference frame (see Section 3.2) to obtain the profiles in the $(x,y)$, $(x,z)$ and $(y,z)$ planes. 

Figure~\ref{fig:proj1} shows the surface density profile (top panel) and line-of-sight velocity dispersion profile (bottom panel) for the $N$-body simulation data (points) and best-fit models (shaded region) on each projection plane. 
The observed differences in the profiles are due to the fact that the $N$-body model shows deviations from the spherical symmetry and the density drops more sharply in the $(y,z)$ plane. Indeed, when looking along the $x$-axis, the Jacobi surface only extends up to $(2/3)r_{\rm J}$ along the $y$-axis and $\sim0.6r_{\rm J}$ along $z$ \citep{Renaud2015}. The non-spherical density distribution of the $N$-body model and the corresponding projection effects also produce a larger velocity dispersion in the outer parts, because the bins outside the Jacobi surface are dominated by PEs. 

As variation in the truncation of the density profiles for different projection angles is predominantly seen beyond $r_{\rm J}$, the best-fit models therefore show little variation finding similar $M_{\rm c}$ and $r_{\rm crit}$ (Table 1). Also in this case, $M_{\rm c}$ and $r_{\rm hm}$ are well reproduced and there is minimal variation in the $\rcrit$ and $\mathcal{F}_{\rm pe}$ therefore the ability of the models to reproduce the global properties of the clusters is not severely affected by projection effects.

\subsection{Clusters on eccentric orbits}

To test how well our equilibrium
models capture the effects induced by an external time-dependent tidal field in the distribution of PEs, we fit them
to $N$-body data of a cluster on an eccentric orbit. Here we take a simulation from C17, with $\lambda=1$ and eccentricity of the clusters orbit of  $\epsilon=0.5$ where $\epsilon = (R_{\rm apo} - R_{\rm peri})/(R_{\rm apo} + R_{\rm peri})$, where $R_{\rm peri}$ and $R_{\rm apo}$ are the perigalactic and apogalactic distance, respectively. We consider three snapshots when the mass first reached approximately 0.4. The snapshots are at pericentre, apocentre and at the position in the orbit equidistant from these two. Figure~\ref{fig:eccen} shows the velocity dispersion and density profiles for the data and the best-fit model for each snapshot. The recovered $r_{\rm  crit}$ of the model is similar for all snapshots, showing that although the model will underpredict and overpredict $r_{\rm J}$ at apocentre and pericentre respectively, it is a good fit to the time-averaged behaviour over one orbit. 
This confirms a finding by \citet{Kupper2010}, who used parametric
fits to the density profiles of $N$-body models on an elliptical orbit.
In this way they recovered an edge radius, and found that it was
nearly constant along the orbit.  In turn this may help to explain a
result of \citet{Cai2016}, namely that the evolution of a
cluster on an eccentric orbit can be approximated by that of a cluster
on a circular orbit with the same dissolution time, if the radius of
orbit is chosen suitably (for modest eccentricities, roughly midway)
between the apo- and pericentric distances of the elliptical orbit.

\section{Observational data}

\begin{figure}
	\centering
	\includegraphics[width=\columnwidth]{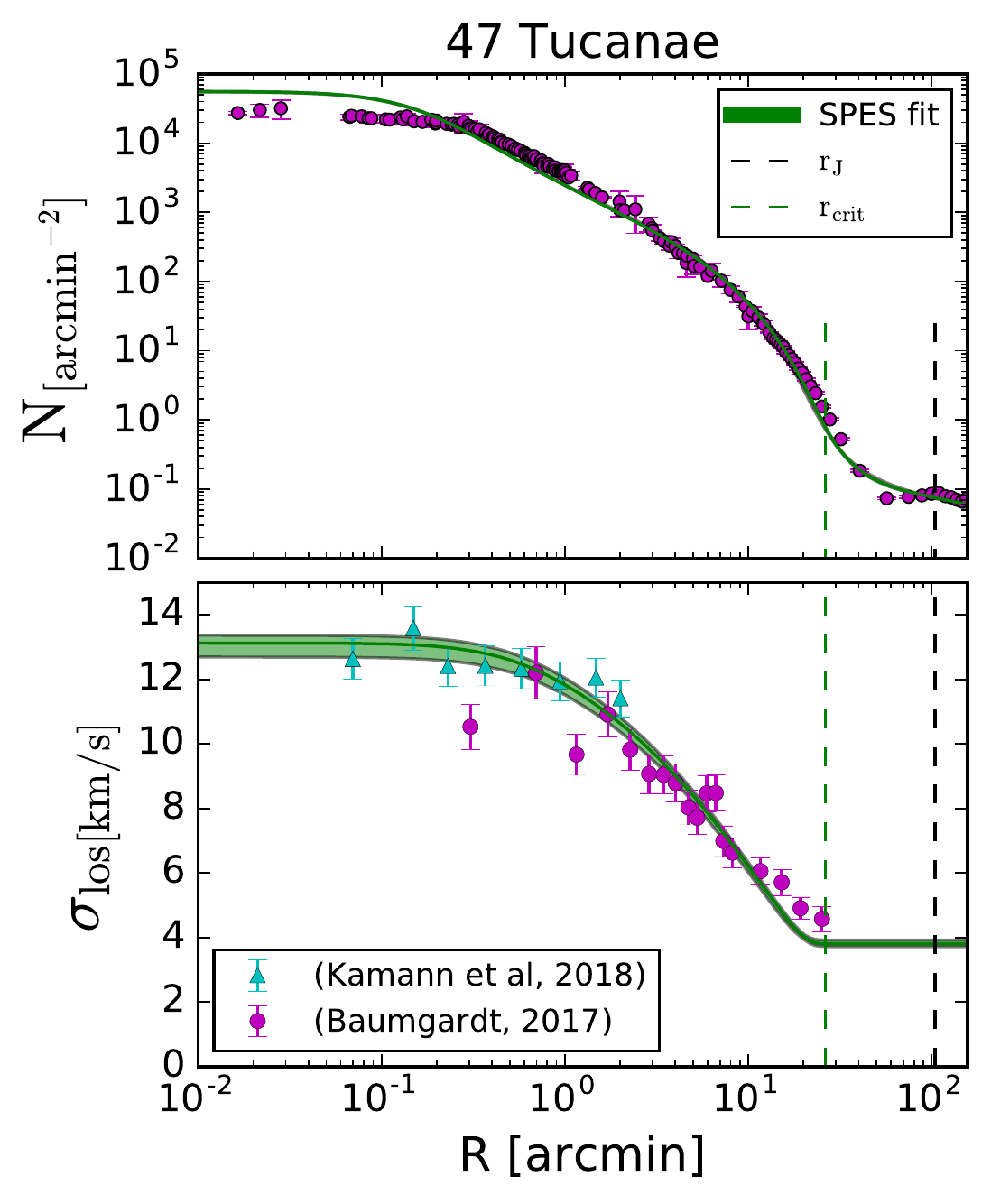}
	\caption{Best-fit {\sc spes} (green) and {\sc limepy} (grey) models compared to the number density (top panel) and line-of-sight velocity dispersion (bottom panel) profiles of 47 Tuc. Data are from \citet{Trager1995}, \citet{Baumgardt2017} and \citet{Kamann2018}. The vertical dashed line marks the position of the Jacobi radius of the cluster, as estimated from  its orbital parameters.} 
	\label{fig:obs1}
\end{figure}

To provide a test of how the {\sc spes} models perform when compared to observational data, we also conduct a preliminary comparison to the number density and the velocity dispersion profiles of the globular cluster 47 Tucanae (47 Tuc).
The choice of 47 Tuc is motivated by the fact that it is a well-known example of a cluster with kinematics that is inconsistent with existing dynamical models considered  \citep{Lane2012}.

In this comparison, we construct a number density profile by combining the surface brightness profile from \citet{Trager1995} and number density profile from the second data release of the {\it Gaia} mission, which are presented in \citet{deBoer2019}. We also fit on the the line-of-sight velocity dispersion profiles ($\sigma_{\rm LOS}$, in km$\,$s$^{-1}$) from \citet{Baumgardt2017} and \citet{Kamann2018}. 

Figure~\ref{fig:obs1} shows the best-fit {\sc spes} model (green shaded region) compared to the surface brightness (top panel) and line-of-sight velocity dispersion (bottom panel) of 47 Tuc. The best-fit model parameters of the fit are $\hat{\phi}_0=9.3$, $\eta=0.30$, $B=0.88$ with scale values $M_{\rm c}=7.0\times10^5\,{\rm M}_{\odot}$ and $\rcrit=11.5$\,arcmin. The best-fit {\sc spes} model reproduces the velocity dispersion profile well, but 
underestimates the last three bins of data. This is also seen in the number density profile, where the model overestimates the central value, and is not able to reproduce the exact shape in the outskirts. This leads to the model underestimating the $r_{\rm J}$ and overestimating the mass when compared to estimates from the Harris catalogue \citep{Harris1996}.  The best-fit $\mathcal{F}_{\rm pe} = 0.038$, but as shown in Section~\ref{Sect:fractionPE}, this value can underestimate the fraction of PEs in the cluster by at least $\sim70\%$.

\section{Discussion \& summary}
\label{Sect:Disc}
We have presented a novel way of including energetically unbound stars in dynamical models of GCs. Our prescription, although based on a simple phase space description of a population of PEs, allows a rapid and convenient self-consistent construction of spherically-symmetric equilibria. This modelling effort was motivated by the peculiarities in observational data in the outer regions of GCs, and in particular by the flattening observed in line-of-sight velocity dispersion profiles and in extended surface density profiles. With \textit{Gaia} providing proper motions of stars in the outskirts of GCs, and allowing us to calculate membership likelihoods for these stars, it is paramount that models which include a description of these behaviours are developed. By including the effects of PEs in a self-consistent, distribution function-based model it is then possible to test if PEs are able to explain the observational data or if some alternative theory is needed, such as the presence of dark matter or of deviations from Newtonian gravity. By characterising the dynamics of stars in the outskirts of GCs, it may be possible to discriminate between these scenarios, and to find clues on the formation and evolution of GCs, which in turn can illuminate the formation and evolutionary processes that shape galaxies.

Even though almost the totality of the `lowered isothermal' models currently available in the literature
are not designed to incorporate the 
presence of PEs, nonetheless the behaviour of some of these unbound stars 
is sufficiently well reproduced, albeit with incorrect underlying physics. This happens because these models describe isolated systems and therefore predict a higher critical energy with respect to the case in which the effects of a tidal potential are included. Therefore, when fitting on data from clusters embedded in external tidal fields, the models can include some PEs between $E_{\rm crit} < E_{\rm J} < \phi_{\rm t}$. This means that a non-spherical model of GCs which includes the tidal potential of the host galaxy could have the correct $E_{\rm crit}$ but as it would have no prescription for dealing with PEs, it would need to increase its $\rcrit$ more than the spherical model to include PEs in the fit by increasing $\phi_{\rm t}$.

Here we developed a physically motivated 
DF-based model which includes a prescription for stars above the critical energy. This was achieved by including two constants in the bound part of the distribution function, which allow the model to have a non-zero density at the critical energy. The constants in the bound DF allow the enforcement of continuity and also smoothness across the critical energy and avoid a discontinuity in the mass distribution. 

We showed that the model accurately reproduces the properties of tidally-perturbed $N$-body models of star clusters, which naturally include PEs (we have conducted a comparison with 
selected direct $N$-body models with $N$=16384 equal-mass particles, originally presented in C17). The best-fit {\sc spes}  model is able to reproduce the mass and half-mass radius of the $N$-body cluster model, and matches the density and velocity dispersion profiles well, including the flattening near $r_{\rm J}$, although is not able to account for increasing velocity dispersion profiles. The {\sc spes} model closely reproduces $r_{\rm J}$ of the $N$-body model, even when data out to $2r_{\rm J}$ is included in the fitting process.

The {\sc spes} models presented here have some limitations. Primarily the under-prediction of the fraction of cluster mass in PEs, $\mathcal{F}_{\rm pe}$. This is due to some assumptions that were required in the construction of the model. These include the assumption of sphericity and the absence of the Galactic tidal potential. This causes a large fraction of the unbound stars to actually be accounted for by the bound part of the chosen DF. We also adopted a simple exponential for the functional form of the unbound part of the DF,  which has some implications on the requirement of continuity of the `stitched' DF. Currently, the {\sc spes} models are defined as continuous and smooth by construction; to force higher order derivatives to be continuous would either over-constrain the model or it would require a different functional form for the bound part, requiring a larger number of parameters. 

Adding the galactic tidal potential to the model may allow for a more accurate recovery of $\mathcal{F}_{\rm pe}$, because then $E_{\rm crit}$ can be recovered more accurately. As part of the unbound population stars would no longer be accounted for by the bound part of the model this will motivate the need for an alternative, more accurate functional form of the DF that better fits the behaviour of these unbound stars. This could then allow for the continuity of higher order derivatives across $\hat{E}=0$. The current definition of the model assumes isotropy and does not account for the possibility of bulk motions of the PEs, which have recently been explored by means of $N$-body simulations (C17; \citealt{Tiongco2016b}). To include this additional layer of kinematic complexity would be an important further step towards a fully realistic description of the phase space behaviour of PEs in star clusters and improve the models ability to discriminate between bound stars and PEs. However, a model which does not include the effects of a galactic potential will not be able to recover both  $r_{\rm J}$ and $\mathcal{F}_{\rm pe}$.

Despite these limitations, the {\sc spes} models are an improvement over existing DF-based models which are unable to account for the presence of a population of energetically unbound stars.
As a proof of concept, we presented a preliminary application of the {\sc spes} models to the number density and velocity dispersion profiles of the Galactic globular cluster 47 Tuc. By using the velocity dispersion data from \citet{Baumgardt2017} and \citet{Kamann2018} and surface brightness profile from \citet{Trager1995} combined with recent {\it Gaia} DR2 data. We showed that the model recovers a $M_{\rm c}$, $r_{\rm hm}$ and $M/L$ close to current estimates, although underestimates $r_{\rm J}$.

The cornerstone ESA mission \textit{Gaia} finally enable us to access the phase space structure of the outskirts of several Galactic globular clusters, therefore it will be paramount to have physically accurate models that are able to describe in more detail and more realistically the expected behaviour of the outer parts of GCs, to be able to correctly infer the properties of the stellar clusters. \citet{deBoer2019} recently fit {\sc spes} models to  {\it Gaia} number density profiles of 81 globular clusters and find that they provide a better prescription near $r_{\rm J}$ than {\sc limepy} models. This is a required first step to determine if any further physics will need to be invoked, such as modified gravity theories or dark matter haloes, to explain the observations. This will in turn provide a method for investigating and possibly discriminating between the formation scenarios and evolutionary behaviour of GCs.

\section*{Acknowledgements}
	
MG acknowledges financial support from the Royal Society (University Research Fellowship), ALV from a Marie Sklodowska-Curie Fellowship (MSCA-IF-EF-RI-658088 NESSY) and the Institute for Astronomy at the University of Edinburgh, AZ from the Royal Society Newton International Fellowship Follow-up Funding, and through a ESA Research Fellowship. IC and MG acknowledge the European Research Council (ERC-StG-335936, CLUSTERS), AZ and ALV the Carnegie Trust for the Universities of Scotland (Research Incentive Grant 70467). We are grateful to Sverre Aarseth and Keigo Nitadori for making {\sc nbody6} publicly available. We also thank Mr. Dave Munro of the University of Surrey for hardware and software support.

\bibliographystyle{mn2e}

\label{lastpage}

\end{document}